\documentclass{aastex631}

\usepackage{subfigure}
\usepackage{color}
\usepackage{amsmath}

\usepackage{ulem}

\begin{document}

\title{
Spectral Properties of Irradiated Circumbinary Disks around Binary Black Holes \\ Governed by Hydrogen Opacities Dependent on Temperature and Density
}


\email{saemi.bang@chungbuk.ac.kr, kimi@chungbuk.ac.kr}

\author{Saemi Bang}
\affiliation{Department of Astronomy and Space Science, Chungbuk National University, Republic of Korea}

\author{Atsuo T. Okazaki}
\affiliation{
Center for Development Policy Studies, 
Hokkai-Gakuen University, Toyohira-ku, Sapporo 062-8605 Japan}

\author{Kimitake Hayasaki}
\affiliation{Department of Astronomy and Space Science, Chungbuk National University, Republic of Korea}
\affiliation{Department of Physical Sciences, Aoyama Gakuin University, Sagamihara 252-5258, Japan}

\begin{abstract}
We study the thermal and spectral properties of irradiated circumbinary disks (CBDs) around binary black holes (BBHs), using analytic, hydrogen-based opacity models that capture dependencies on temperature, density, and ionization. We solve the vertical hydrostatic equilibrium and energy balance, assuming gas pressure only, using Rosseland-mean opacities from free–free and bound–free absorption plus electron scattering, with ionization fractions given by the Saha equation.
Four opacity models are considered, including a reference model with no physical opacity, constructed by \citet{2024ApJ...975..141L}, and three physically motivated alternatives. The midplane temperature profiles show significant variation across models, while the surface temperature remains largely unchanged in regions dominated by viscous heating.
Opacity effects become pronounced in the outer disk, where irradiation reprocessing shapes the IR-optical continuum. Bound-free opacity introduces flattening and a mid-frequency peak in the spectral energy distribution. We compute spectra of a triple disk system including the CBD and two accreting minidisks. The high-frequency peak arises from the hot minidisks, while the low-frequency excess originates from irradiated outer CBD layers. Comparing model spectra with detection limits of Subaru, JWST, and Swift, we find that BBH systems within $\sim10$ Mpc can exhibit a detectable IR excess. Our results highlight the need for physically consistent opacity modeling to interpret electromagnetic (EM) signatures of BBHs approaching coalescence and support integration of metallicity-dependent opacity tables. Our opacity-informed framework for irradiated CBDs provides an EM template for identifying stellar- to intermediate-mass BBHs in a mass range sparsely sampled by LISA, thereby bridging the gravitational-wave--EM gap with testable IR/optical signatures.
\end{abstract}


%
\section{Introduction} 
\label{sec:intro}
%

Stellar-mass binary black holes (BBHs), composed of two black holes with individual masses between $3\,M_\odot$ and $100\,M_\odot$, can form either through the isolated binary evolution of massive stars \citep{2016A&A...588A..50M} or through dynamical interactions in dense stellar systems such as globular clusters \citep{2015ApJ...800....9M}. Since the first direct detection of a black hole merger by the LIGO Observatory (event GW150914) \citep{2016PhRvL.116f1102A}, a growing number of such events have been reported \citep{2023PhRvX..13a1048A,2023ApJ...946...59N}, including those suggesting the existence of intermediate-mass black holes (IMBHs). On a larger scale, supermassive black hole (SMBH) binaries are thought to reside at the centers of active galactic nuclei (AGNs) and may play a key role in galaxy evolution as well as structure formation in the early universe. At all these mass scales, the presence of circumbinary disks (CBDs) around black-hole binaries provides an important way to probe binary evolution via electromagnetic (EM) emission complementary to gravitational waves (GWs).

The thermal and spectral properties of a CBD depend critically on its internal heating and radiative cooling processes. In particular, CBDs are expected to be heated by viscous dissipation and, in some cases, by irradiation from an accreting minidisk around each black hole. A physically consistent treatment of irradiation is therefore essential for predicting CBD spectra and EM observables.

In this context, the single–black-hole disk literature provides well-tested baselines for irradiation physics. Early studies established the framework of X-ray–heated, reprocessing disks \citep{hayakawa_x-ray_1981}, and general-relativistic “returning radiation” can self-irradiate the outer disk around a Schwarzschild black hole \citep{sanbuichi_self-irradiated_1993}, building on earlier Newtonian treatments of flared, self-irradiated disks \citep{fukue_self-irradiated_1992}. In stellar-mass X-ray binaries, these ideas matured into quantitative reprocessing models that link optical/infrared (IR) light to the X-ray luminosity and help explain outburst decays and the maintenance of the hot state \citep{vanParadijsMcClintock1994,king_light_1998,Dubus1999,DubusHameuryLasota2001}. As contemporary observational anchors, irradiated disk models have been fit jointly to X-ray and optical data of ultraluminous X-ray sources (e.g. \citealt{2014MNRAS.444.2415S}), and recent multi-wavelength campaigns in single-black-hole systems now recover ultraviolet (UV)/optical lags \citep{Oknyansky2025_NGC2617_RM} and spectral energy distributions (SEDs) consistent with X-ray thermal reprocessing \citep{Yoshitake2024_MAXIJ1820_Rebrightening}. These single-black-hole baselines motivate how we incorporate irradiation from the minidisks in the CBD context below.

A recent study by \citet{2024ApJ...975..141L} investigated the effect of irradiation from such minidisks in shaping the CBD temperature profile and spectrum, showing that reprocessed emission can lead to a characteristic IR peak in the SED. However, their analysis assumed the disk surface temperature equals the disk midplane temperature, equivalent to adopting a fixed effective optical depth $\tau=16/3$, thereby neglecting the strong temperature and ionization dependence of opacity in cooler, denser regions of the CBD.

In reality, the opacity of a hydrogen-rich disk varies significantly with local temperature $T$, density $\rho$, and ionization degree $x_e(T)$ \citep{rybicki_radiative_1979}. In particular, for the midplane temperature of an accretion disk in the range $10^3~{\rm K} \lesssim T \lesssim 10^4~{\rm K}$, the dominant sources of opacity are known to be bound-free ($\kappa_{\rm bf}$) and free-free ($\kappa_{\rm ff}$) absorption, while electron scattering ($\kappa_{\rm es}$) typically dominates for $T\gtrsim10^4\, {\rm K}$ (e.g., \citealt{1973A&A....24..337S, kato_black-hole_2008}). These absorption opacities are highly sensitive to $T$ and $x_e$, the latter of which is determined by the Saha equation \citep{Saha01101920}.

In BBH systems, ionizing UV photons and X-rays emitted by the minidisks can strongly modify the ionization state of the CBD surface layers, and hence the relevant opacities. In the irradiation-dominated surface layers of a CBD, such hard photons can suppress dust and reduce the importance of metal-line absorption by driving the gas toward high ionization \citep{Barvainis1987}. In the irradiation-dominated layers relevant for CBD reprocessing, the opacity can therefore be treated as effectively hydrogen-dominated, together with electron scattering. Motivated by this, and to isolate the irradiation/ionization physics relevant to CBD reprocessing, we adopt a clean baseline model with a pure-hydrogen composition: we use analytic Kramers-type prescriptions for $\kappa_{\rm ff}$ and $\kappa_{\rm bf}$ together with $\kappa_{\rm es}$, and compute the ionization fraction self-consistently from the Saha equation. This baseline allows us to quantify how the opacity prescription controls (i) the midplane--surface temperature contrast and (ii) the resulting IR/optical emission from the outer CBD. We revisit the role of metals, dust, and tabulated opacities in Section~\ref{sec:5.1}.

In this paper, we examine the impact of different opacity sources on the thermal structure and the spectrum of an irradiated CBD. We focus on three opacity prescriptions: $\kappa_{\rm es}$, $\sqrt{\kappa_{\rm es}\kappa_{\rm ff}}$, and $\sqrt{\kappa_{\rm a}(\kappa_{\rm a} + \kappa_{\rm es})}$, where $\kappa_{\rm a}=\kappa_{\rm ff}+\kappa_{\rm bf}$, and assess how they affect the radial profiles of the CBD aspect ratio $H/r$, the CBD midplane temperature $T_{\rm c}$, the surface temperature $T_{\rm s}$, and the spectrum emanating from the CBD surface.

For this purpose, we use Rosseland-mean opacities computed from analytic Kramers' formulas for hydrogen bound-free and free-free absorption processes, incorporating the ionization degree calculated from the Saha equation. We solve the energy balance equation, 
\begin{equation} 
Q_{\rm vis} + Q_{\rm irr} + Q_{\rm tid} = Q_{\rm rad},
\label{eq:eq1}
\end{equation} 
where $Q_{\rm vis}$ is the viscous heating rate, $Q_{\rm irr}$ is the irradiation heating rate, $Q_{\rm tid}$ is the heating rate due to dissipation of tidally excited density waves near the cavity rim, and $Q_{\rm rad}$ is the radiative cooling rate depending on $\kappa(\rho,T_{\rm c},x_e(T_{\rm c}))$.

Throughout this paper, we adopt a one-dimensional (1D), azimuthally and orbit-averaged steady framework\footnote{This time-averaged treatment is justified because the viscous timescale at the inner edge, $\tau_{\rm vis}(r_{\rm in})$, is much longer than the binary orbital period $P_{\rm orb}$, as quantified in equation (\ref{eq:vistime}).}. In all CBD models presented in this paper we set $Q_{\rm tid}=0$. Analytic studies show that the dissipation of these waves is confined to a narrow annulus of width $\sim$ a few scale heights $H$ at $r\simeq r_{\rm in}$ \citep{2009MNRAS.398.1392L,2013ApJ...774..144R}. For the geometrically thin disks considered here, with $H/r \lesssim 10^{-2}$, this tidally heated region occupies only a small fraction $\sim H/r \ll 1$ of the CBD surface and thus contributes negligibly to the disk-wide energy budget and to the outer-disk surface temperature that sets the global SED. This choice is also consistent with our surface-density prescription, which ignores tidal torques away from the rim (see equations~(\ref{eq:sden_orig}) and (\ref{eq:sden}) of Section~\ref{subsec:tau} for details). The resulting models allow us to quantify how microphysical opacity prescriptions shape the thermal state and spectral signature of the CBD with and without irradiation. 

Our study aims to establish a physically consistent baseline for understanding irradiated CBDs in a low temperature regime and to facilitate future comparisons with models incorporating metallicity, dust, or full opacity tables. These results may also aid in the interpretation of EM counterparts to black hole mergers or merging BHs detected via GWs.

In Section~\ref{sec:model}, we present the basic framework for modeling a 1D, geometrically thin, optically thick, irradiated CBD with the detailed description of the three opacity prescriptions. In Section~\ref{sec:method}, we numerically solve the energy equation of the CBD. Among the three opacity models, the model involving combined free-free absorption, bound-free absorption, and electron scattering processes requires numerical integration due to their nonlinear dependence on the density, temperature, and ionization degree. In Section~\ref{sec:results}, we present the resulting radial profiles of disk temperature, the emergent spectra from the CBD, and the combined spectra from the CBD and minidisks, referred to as the triple disk spectra. Sections~\ref{sec:dis} and~\ref{sec:conclusion} are devoted to discussion and conclusions, respectively.

\begin{figure*}[htbp]
    \centering    \includegraphics[scale=0.9]{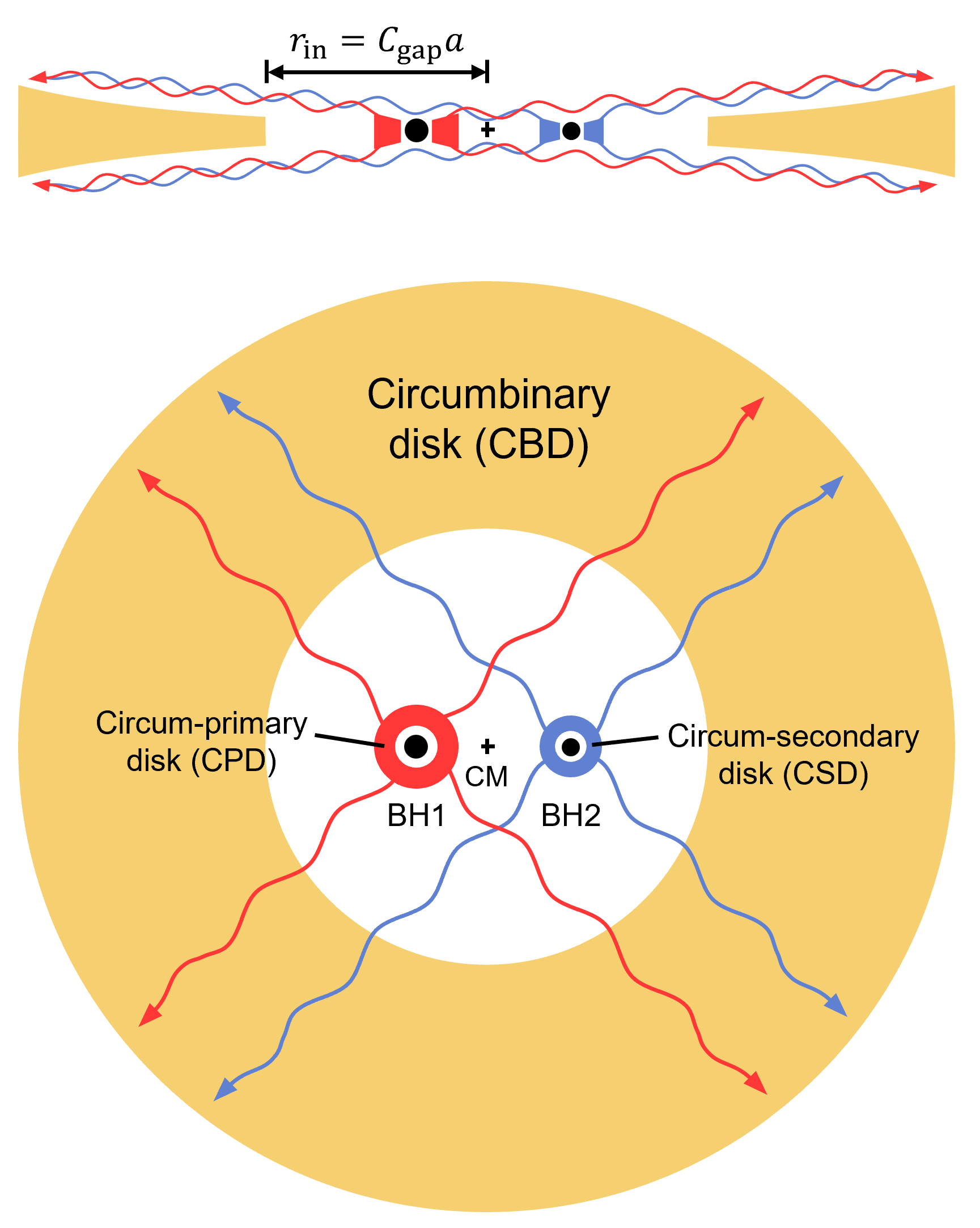}
    \caption{
    Schematic illustration of an irradiated circumbinary disk (CBD) surrounding two minidisks orbiting the primary and secondary black holes. The minidisks, referred to as the circum-primary disk (CPD) and the circum-secondary disk (CSD), are gravitationally bound to the primary (BH1) and secondary (BH2) black holes, respectively. The yellow ochre, blue, and red shaded regions represent the CBD, CPD, and CSD, respectively. The two black circles at the center denote the black holes, with their common center of mass (CM) marked by a plus sign. The blue and red wavy arrows indicate the trajectories of irradiating photons emitted from the minidisks. The inner edge of the CBD is located at a radius $r_{\rm in} = C_{\rm gap} a$, where $C_{\rm gap}$ is a parameter characterizing the size of the gap between the CBD and the binary, and $a$ is the binary's semi-major axis.
    }
    \label{fig:tripledisk}
\end{figure*}
%
\section{Model} 
\label{sec:model}
%

We consider a 1D, axisymmetric, and steady-state CBD surrounding a BBH system, computed from azimuthally and orbit-averaged equations that neglect explicitly time-dependent and non-axisymmetric components of the binary tidal potential. The CBD is assumed to be optically thick and geometrically thin, and to evolve under viscous dissipation in the presence of external irradiation from the minidisks. In this framework, the binary affects the CBD structure only through the geometric truncation at the cavity edge described by equation (\ref{eq:rin}) below and through the irradiation source term defined by equation~(\ref{eq:qirr}) in Section~\ref{sec:3.2}

Our goal is to investigate how the thermal structure of the CBD and its emergent spectrum respond to different physical forms of the Rosseland mean opacity, particularly in the midplane temperature range $10^3\,{\rm K}\lesssim T_{\rm c} \lesssim 10^5\,{\rm K}$, where temperature- and ionization-dependent hydrogen absorption dominates. To isolate this dependence, we assume a pure hydrogen composition.

The total mass of the binary is defined as $M = M_1 + M_2$, where $M_1$ and $M_2$ are the masses of the primary and secondary black holes, respectively. The binary follows a circular orbit with a semi-major axis $a = a_0\, r_{\rm S}$, where $a_0$ is a dimensionless scaling factor and the Schwarzschild radius, $r_{\rm S}$, is defined as
\begin{eqnarray}
    r_{\rm S}
    &&
    =\frac{2GM}{c^2}
    \nonumber \\
    &&
    \sim3.0\times10^{7}\,{\rm cm}\left(\frac{M}{100\,M_\odot}\right),
    \label{eq:rs}
\end{eqnarray}
where $G$ is the gravitational constant and $c$ is the speed of light. As illustrated in Figure~\ref{fig:tripledisk}, the CBD inner-edge radius is given by 
\begin{eqnarray}
    r_{\rm in}
    &&
    =C_{\rm gap}a
    \nonumber \\
    &&
     =6.0\times10^{10}\,{\rm cm}
     \left(\frac{C_{\rm gap}}{2}\right)
     \left(\frac{a_0}{1000}\right)
     \left(\frac{M}{100\,M_\odot}\right)
     \label{eq:rin}
\end{eqnarray}
with $1.6\lesssim C_{\rm gap}\lesssim4$ \citep{artymowicz_dynamics_1994,2023ApJ...949L..30D}. We adopt $C_{\rm gap}=2$ as a fiducial value throughout this paper. The binary enters explicitly through the geometric truncation of the CBD at the cavity edge, $r_{\rm in}$. This prescription suppresses local viscous heating for $r<r_{\rm in}$ and limits the CBD's emitted area and the angles at which the CBD is irradiated. Apart from this truncation radius, our models do not consider time-dependent, non-axisymmetric tidal heating or inner torque terms.

It is noted that irradiation originating near the inner edge of each minidisk is unlikely to affect the spectrum of the outer region, since the radius at which irradiation heating dominates over viscous heating is well beyond the disk outer edge of each minidisk. In addition, photons emitted from the inner edge of the CBD are primarily in the optical band and have significantly lower energies than those from the inner edges of the minidisks. This difference arises because the temperatures of the minidisks are higher than the CBD temperature, following the standard disk relation $T \propto r^{-3/4}$ \citep{1981ARA&A..19..137P}. Consequently, the contribution of irradiation from the CBD’s inner edge to the overall SED is negligible. We therefore omit the irradiation heating rate from the CBD inner edge in the energy equation.

The radiative cooling rate from both sides of the disk surface is given by
\begin{eqnarray} 
Q_{\rm rad} = 2 \sigma T_{\rm s}^4, 
\label{eq:qrad} 
\end{eqnarray}
where $T_{\rm s}$ is the surface temperature and $\sigma$ is the Stefan-Boltzmann constant.
Assuming local radiative equilibrium in the vertical direction of the CBD, the radiative flux can also be expressed as
\begin{eqnarray} 
F_{\rm rad} = \frac{32}{3} \frac{\sigma T_{\rm c}^4}{\tau}, 
\label{eq:frad} 
\end{eqnarray}
where $T_{\rm c}$ is the midplane temperature and 
\begin{eqnarray}
    \tau 
    = 
    \frac{1}{2}\kappa(\rho, T_{\rm c}, x_e) \Sigma
\end{eqnarray}
is the optical depth, with $\Sigma$ being the surface density of the disk. In our formulation, the opacity $\kappa(\rho, T_{\rm c}, x_e)$ consists of a sum of contributions from electron scattering, free-free absorption, and bound-free absorption (see Section~\ref{sec:method} for details). Equating equation~(\ref{eq:qrad}) with equation~(\ref{eq:frad}) yields the relation between the surface and midplane temperatures: 
\begin{eqnarray}
    T_{\rm s}=\frac{2}{3^{1/4}}\frac{1}{\tau^{1/4}}T_{\rm c}.
    \label{eq:tstc}
\end{eqnarray}

According to the standard disk theory, the heating rate due to the viscous heating of the disk is given by 
\begin{eqnarray}
    Q_{\rm vis}
    = 
    \frac{3GM\dot M}{4\pi r^3}
    \label{eq:qvis}
\end{eqnarray}
for $r\gg{r_{\rm ISCO}}$, where $r_{\rm ISCO}=3\,r_{\rm S}$ is the radius at the innermost stable circular orbit (ISCO) around a Schwarzschild black hole of mass $M$. We parameterize $\dot{M}$ in terms of the Eddington accretion rate as 
\begin{eqnarray}
\dot{M}=\dot{m}\frac{L_{\rm Edd}}{c^2},
\label{eq:mdot}
\end{eqnarray} 
 where $\dot{m}$ denotes the dimensionless accretion rate, which is set to unity throughout this paper, and 
\begin{eqnarray}
     L_{\rm Edd} = \frac{4\pi GMc}{\kappa_{\rm es}}
     \label{eq:ledd}
\end{eqnarray}
 is the electron-scattering based Eddington luminosity. The electron-scattering opacity is set to be $\kappa_{\rm es} = 0.2(1 + X)\,{\rm cm^2\,g^{-1}}$, where $X$ is the hydrogen mass fraction. In the following we assume $X=1$ and thus $\kappa_{\rm es}=0.4\,{\rm cm^2\,g^{-1}}$.

The disk viscosity is given by \citep{1973A&A....24..337S}
\begin{eqnarray}
    \nu= \frac{2}{3}\alpha_{\rm SS}c_{\rm s}H
    \label{eq:nu}
\end{eqnarray}
with the viscosity parameter $\alpha_{\rm SS}$, where $c_{\rm s}$ is the sound speed. Assuming that the CBD is vertically supported by gas pressure and is in hydrostatic equilibrium for the direction perpendicular to the disk plane, the sound speed $c_{\rm s}$ is given by
\begin{eqnarray}
    c_{\rm s}=\Omega{H},
    \label{eq:hydrostaticeq}
\end{eqnarray}
where $\Omega=\sqrt{GM/r^3}$ is the Keplerian angular frequency of the CBD.

The ratio between the CBD viscous timescale $\tau_{\rm vis}(r)=r^2/\nu$ at the CBD inner edge and the binary orbital period $P_{\rm orb}$ is given by using equations~(\ref{eq:nu}) and (\ref{eq:hydrostaticeq}) as
\begin{eqnarray}  
     \frac{\tau_{\rm vis}\left(r_{\rm in}\right)}{P_{\rm orb}}
     &\sim&
     2.0\times10^4\left(\frac{C_{\rm gap}}{2}\right)^3\left(\frac{a_{0}}{1000}\right)^{3/2}\left(\frac{\alpha_{\rm SS}}{0.1}\right)^3
     \nonumber \\
     &\times&
     \left(\frac{H/r}{0.01}\right)^2,
     \label{eq:vistime}
\end{eqnarray}
where $P_{\rm orb}=2\pi/\Omega\sim 2.8\times10^2 \, {\rm s} \, \left(a_{0}/1000\right)^{3/2}
\left(M/100M_\odot\right)^{-1/2}$.

Since equation~(\ref{eq:vistime}) suggests that 
$\tau_{\rm vis}\gg P_{\rm orb}$, we assume that the binary system with triple disk composed of CBD, CPD, and CSD is in a quasi-steady state. This allows us to impose on the following relation on the accretion rates between CBD, CPD, and CSD: 
\begin{eqnarray}
 \dot{M}=\dot{M}_1+\dot{M}_2,
\end{eqnarray}
where $\dot{M}_1$ and $\dot{M}_2$ are the CPD and CSD accretion rates, respectively. Also, we assume that the mass accretion rate ratio is equal to the binary mass ratio $q$, i.e. $\dot{M}_2/\dot{M}_1=q$. These relations provide
\begin{eqnarray}
\dot{M}_1= \frac{1}{1+q}\dot{M},\,\,
\dot{M}_2=\frac{q}{1+q}\dot{M}.
\label{eq:mdots12}
\end{eqnarray}
In the equal-mass limit $q=1$, symmetry implies that the instantaneous accretion rates onto the two minidisks are comparable, $\dot{M}_2/\dot{M}_1 \simeq q$, as found in hydrodynamic simulations of circumbinary accretion \citep{2014ApJ...783..134F,DuffellEtAl2020}. Because our study focuses on this limit and is motivated by these findings, we adopt the prescription of $\dot{M}_2/\dot{M}_1=q$ throughout this paper.

The bolometric luminosities of the primary and secondary black holes are given by 
\begin{equation}
    L_1=\frac{1}{6}\dot{M}_1c^2,\,\,
    L_2=\frac{1}{6}\dot{M}_2c^2.
    \label{eq:bol_lumi}
\end{equation}

The equation of state for an ideal gas gives the sound speed as
\begin{eqnarray}
c_{\rm s}=\sqrt{\frac{R_{\rm g}}{\mu}T_{\rm c}},
\label{eq:soundspeed}
\end{eqnarray}
where $R_{\rm g}$ is the gas constant and $\mu$ 
is the mean molecular weight. For hydrogen gas, $\mu$ is given by 
\begin{eqnarray}
    \mu=\frac{1}{1+x_e(T_{\rm c})},
    \label{eq:mu}
\end{eqnarray}
where $x_e(T_{\rm c})$ denotes the ionization degree, where $x_e = 0$ and $x_e = 1$ correspond to neutral hydrogen atoms and fully ionized hydrogen gas, respectively. The ionization fraction is determined by the Saha equation \citep{Saha01101920}:
\begin{eqnarray}
    \frac{x_e^2}{1 - x_e} 
    =
    2
    \left(
    \frac{2 \pi m_e k T_{\rm c}}{h^2}
    \right)^{3/2}
    \frac{1}{n_{\rm H}}
    \frac{g_{\rm i}}{g_{\rm n}}
    \exp\left( -\frac{\chi}{k T_{\rm c}} \right),
    \label{eq:saha}
\end{eqnarray}
where $m_e$ is the electron mass, $h$ is the Planck constant, $n_{\rm H}$ is the number density of hydrogen atoms, $k$ is the Boltzmann constant, $g_{\rm i}$ and $g_{\rm n}$ are the statistical weights of the ionized and neutral states, respectively, and $\chi$ is the ionization energy. Throughout this paper, we adopt $g_{\rm i} = 1$, $g_{\rm n} = 1$, and $\chi = 13.6\,{\rm eV}$.

Equating equation~(\ref{eq:hydrostaticeq}) with equation~(\ref{eq:soundspeed}) gives the CBD midplane temperature:
\begin{eqnarray}
T_{\rm c}=\frac{1}{R_{\rm g}}
\frac{1}{1+x_e}
\left(\frac{H}{r}\right)^2
\frac{GM}{r}.
\label{eq:mid-plane-temp}
\end{eqnarray}

Throughout this work, the vertical support and sound speed are computed with an ideal gas equation of state, with the mean molecular weight determined by the Saha ionization fraction. Consequently, the pressure that enters the hydrostatic balance and the energy equation is gas pressure alone. Neither the vertical force balance nor the equation of state includes radiation pressure, and magnetic pressure is likewise neglected. This treatment is consistent with our focus on cold, geometrically thin, optically thick CBDs, where $T_{\rm c} \sim 10^3$-$10^5\,{\rm K}$, and irradiation primarily affects the surface layers rather than the midplane structure. We discuss the applicability and limitations of this approximation in Section~\ref{sec:dis}.

%
\subsection{Disk Opacity}
%

In partially ionized hydrogen gas, the dominant sources of opacity are free-free absorption ($\kappa_{\rm ff}$), bound-free absorption ($\kappa_{\rm bf}$), and electron scattering ($\kappa_{\rm es}$). The free-free opacity arises from the bremsstrahlung absorption by electrons interacting with protons, and is given approximately by (e.g., \citealt{rybicki_radiative_1979})
\begin{eqnarray}
    \kappa_{\rm ff} 
    =
    \kappa_{\rm ff,0}
    x_e^2 
    \rho\,T_{\rm c}^{-7/2}
    \,{\rm cm^2 g^{-1}}
    \label{eq:kff}
    ,
\end{eqnarray}
where $\kappa_{\rm ff,0}=1.1 \times 10^{23}\,{\rm cm^5 g^{-2}K^{7/2}}$ and $\rho$ is the mass density in g\,cm$^{-3}$. The bound-free opacity, caused by photoionization of neutral hydrogen atoms, is typically written as 
\begin{eqnarray}
\kappa_{\rm bf}
=
\kappa_{\rm bf,0}
(1 - x_e)\rho\,T_{\rm c}^{-7/2}
\,{\rm cm^2 g^{-1}}
\label{eq:kbf}
,
\end{eqnarray}
where $\kappa_{\rm bf,0}=4.3\times10^{25}\,{\rm cm^5 g^{-2}K^{7/2}}$ (e.g., \citealt{rybicki_radiative_1979}).
While $\kappa_{\rm bf}$ is most effective in the temperature range $T_{\rm c}\sim5000$-$10^4$ K, it can still be applied down to $T_{\rm c} \sim 1000$-$5000$ K in hydrogen-rich accretion disks.

However, in realistic astrophysical conditions, especially at $T \lesssim 5000$ K, the true opacity deviates significantly from the hydrogen $\kappa_{\rm bf}$ because of contributions from metals and dust grains (e.g., \citealt{2003A&A...410..611S,2005ApJ...623..585F}). These include the line and continuum absorption by molecules and solid particles, which can increase the opacity by orders of magnitude. Despite this, using $\kappa_{\rm bf}$ derived from pure hydrogen models remains useful for highlighting the role of hydrogen ionization and providing an upper bound on the opacity in irradiated disks. This approach is particularly appropriate for idealized accretion disks composed of hydrogen, enabling us to isolate fundamental physical effects without introducing chemical complexity.

The effective opacities used in our model are given by \citep{rybicki_radiative_1979}
\begin{eqnarray}
\kappa_1
&=&
\kappa_{\rm es},
\label{eq:k1}
\\
\kappa_2
&=&
\sqrt{\kappa_{\rm es}\kappa_{\rm ff}},
\label{eq:k2}
\\
\kappa_3
&=&
\sqrt{\kappa_{\rm a}\left(\kappa_{\rm a}+\kappa_{\rm es}\right)},
\label{eq:k3}
\end{eqnarray}
where the absorption opacity $\kappa_{\rm a}$ is given by
\begin{equation}
\kappa_{\rm a}=\kappa_{\rm ff}+\kappa_{\rm bf}.
\label{eq:ka}
\end{equation}

To explore how different opacity prescriptions affect the thermal structure and emergent spectrum of the CBD, we consider four representative models. Model 0 reproduces the fiducial case of \cite{2024ApJ...975..141L}, in which no physical source of opacity is included by assuming that $T_{\rm s} = T_{\rm c}$. This model serves as a baseline for comparison with the three physically motivated opacity models. Model~1 includes only electron scattering opacity, representing the fully ionized regime where scattering dominates. Model~2 assumes a hybrid regime where electron scattering is stronger than but not entirely dominant over free-free absorption. Model~3 employs a more integrated treatment by accounting for both the two absorption processes and electron scattering. Table~\ref{tbl:opmodels} summarizes the adopted opacity prescriptions, associated optical depth expressions, and boundary conditions for each model.

%
\section{Method} 
\label{sec:method}
%

Solving the energy equation yields the disk aspect ratio $H/r$, which can then be substituted into the hydrostatic equation to finally obtain the radial distribution of the CBD temperature. In the following, we describe the method to solve the energy equation numerically. First, we introduce the following dimensionless variables:
\begin{eqnarray}
    \xi&\equiv&\frac{r}{r_{\rm in}}
    \hspace{2.5mm}{\rm and}\hspace{2.5mm}
    Y\equiv\frac{H}{r}.
    \label{eq:dimlessparams}
\end{eqnarray}

Equation~(\ref{eq:mid-plane-temp}) is rewritten with equation~(\ref{eq:dimlessparams}) as
\begin{equation}
    T_{\rm c}(1+x_{e}(T_c))=T_0\frac{Y^2}{\xi},
    \label{eq:tct0}
\end{equation}
where we define the dynamical temperature at $r_{\rm in}$ as 
\begin{eqnarray}
    T_0
    &&
    =
    \frac{1}{R_g}\frac{GM}{r_{\rm in}}
    \sim 2.7\times10^9\,{\rm K}
    \left(\frac{C_{\rm gap}}{2}\right)^{-1}
    \left(\frac{a_0}{1000}\right)^{-1}.
    \nonumber 
    \label{eq:t0}
\end{eqnarray}

%
\subsection{Optical depth prescriptions}
\label{subsec:tau}
%
Following \citet{2013ApJ...774..144R}, the conservation law of the radial angular momentum flux $F_J$ in a steady state is given by $F_J=3\pi \nu\Sigma{l} - \dot{M}l$, where $l=r^2\Omega$ and a constant inner-boundary contribution $F_{J}$ that encapsulates the binary torque. We then obtain the surface density profile
\begin{equation}
\Sigma
=
\Sigma_0
\Bigl[
1+\epsilon_{\rm bin}
\xi^{-1/2}
\Bigr],
\label{eq:sden_orig}
\end{equation}
where $\epsilon_{\rm bin} = F_{J}/(\dot{M}\sqrt{GMr_{\rm in}})$ is the binary-torque parameter, and $\Sigma_0$ is the surface density of the standard disk model: 
\begin{eqnarray}
    \Sigma_0
    &=&
    \frac{\dot{M}}{3\pi\nu}
    =
    \sqrt{2}\frac{\dot{m}}{\alpha_{\rm SS}}
    \frac{1}{\sqrt{C_{\rm gap}a_0}}
    \frac{1}{\kappa_{\rm es}}
    \frac{1}{Y^2\xi^{1/2}},
    \label{eq:sden}
\end{eqnarray}
Here, equations (\ref{eq:rin}), (\ref{eq:mdot}), (\ref{eq:nu}), (\ref{eq:hydrostaticeq}), and (\ref{eq:dimlessparams}) are used for the derivation. Equation~(\ref{eq:sden}) makes explicit how a nonzero inner angular momentum flux inherited from the binary ($\epsilon_{\rm bin}\neq 0$) produces an inner pile-up ($\epsilon_{\rm bin}>0$) or depletion ($\epsilon_{\rm bin}<0$)\footnote{In the circumbinary systems of interest, the binary is expected to lose angular momentum to the CBD, so that the corresponding inner torque and $F_J$ are non-negative ($\epsilon_{\rm bin}\ge0$); we therefore do not consider the depletion regime with $\epsilon_{\rm bin}<0$ in this work.}, while reducing to $\Sigma_0$ at the zero-torque limit ($\epsilon_{\rm bin}=0$). For simplicity, we take $\epsilon_{\rm bin}=0$ throughout this paper.

The mass density is then given by
\begin{eqnarray}
\rho=\frac{1}{2}\frac{\Sigma}{H}
&=&
\frac{1}{2}\frac{r}{H}\frac{\Sigma}{r_{\rm S}}\frac{r_{\rm S}}{r_{\rm in}}\frac{r_{\rm in}}{r}
=
\frac{1}{2}\frac{1}{Y\xi}\frac{\Sigma}{r_{\rm S}}\frac{1}{C_{\rm gap}a_0}\
\nonumber \\
&=&
\frac{1}{\sqrt{2}}
    \frac{\dot{m}}{\alpha_{\rm SS}}
    \frac{1}{(C_{\rm gap}a_0)^{3/2}}
\frac{1}{r_{\rm S}\kappa_{\rm es}}
 \frac{1}{Y^3\xi^{3/2}} .
\label{eq:rho}
\end{eqnarray}
Substituting equations~(\ref{eq:tct0}) and~(\ref{eq:rho}) into equations~(\ref{eq:kff}) and~(\ref{eq:kbf}) yields the free-free and bound-free absorption opacities in the form
\begin{eqnarray}
\kappa_{\rm ff}
&=&
\kappa_{\rm ff,0}\,\rho\,T_{\rm c}^{-7/2}
\nonumber \\
&=&
\frac{1}{\sqrt{2}}
\frac{\dot{m}}{\alpha_{\rm SS}}
\frac{1}{(C_{\rm gap}a_0)^{3/2}}
\left[
\frac{\kappa_{\rm ff,0}}{\kappa_{\rm es}}
\frac{x_e^2}{r_{\rm S}}
\left(\frac{1+x_e}{T_0}\right)^{7/2}
\right]
\frac{\xi^2}{Y^{10}},
\nonumber \\
\label{eq:kff2}
\\
\kappa_{\rm bf}
&=&
\kappa_{\rm bf,0}\,\rho\,T_{\rm c}^{-7/2}
\nonumber \\
&=&
\frac{1}{\sqrt{2}}
\frac{\dot{m}}{\alpha_{\rm SS}}
\frac{1}{(C_{\rm gap}a_0)^{3/2}}
\left[
\frac{\kappa_{\rm bf,0}}{\kappa_{\rm es}}
\frac{1-x_e}{r_{\rm S}}
\left(\frac{1+x_e}{T_0}\right)^{7/2}
\right]
\frac{\xi^2}{Y^{10}}.
\nonumber 
\label{eq:kbf2}
\\
\end{eqnarray}
Equation~(\ref{eq:ka}) can be rewritten as
\begin{eqnarray}
\kappa_{\rm a}
&=&
\frac{1}{\sqrt{2}}
\frac{\dot{m}}{\alpha_{\rm SS}}
\frac{1}{(C_{\rm gap}a_0)^{3/2}}
\frac{\kappa_{\rm ff,0}}{\kappa_{\rm es}}
\frac{1}{r_{\rm S}}
\nonumber \\
&\times&
x_e^2
\left(\frac{1+x_e}{T_0}\right)^{7/2}
\left[
1
+
\frac{1-x_e}{x_e^2}
\frac{\kappa_{\rm bf,0}}{\kappa_{\rm ff,0}}
\right]
\frac{\xi^2}{Y^{10}}.
\label{eq:ka2}
\end{eqnarray}
Equations~(\ref{eq:k2}) and (\ref{eq:k3}) are also rewritten as
\begin{eqnarray}
\kappa_2
&=&
\frac{1}{2^{1/4}}
\left(\frac{\dot{m}}{\alpha_{\rm SS}}\right)^{1/2}
\frac{1}{(C_{\rm gap}a_0)^{3/4}}
\left(\frac{\kappa_{\rm ff,0}}{r_{\rm S}}\right)^{1/2}
\nonumber \\
&\times&
x_e
\left(
\frac{1+x_e}{T_0}
\right)^{7/4}
\frac{\xi}{Y^5} ,
\\
\kappa_3
&=&
\frac{1}{\sqrt{2}}
\frac{\dot{m}}{\alpha_{\rm SS}}
\frac{1}{(C_{\rm gap}a_0)^{3/2}}
\frac{\kappa_{\rm ff,0}}{\kappa_{\rm es}}
\frac{1}{r_{\rm S}}
\left(\frac{1+x_e}{T_0}\right)^{7/2}
\nonumber \\
&\times&
\left[
x_e^2
+
(1-x_e)
\frac{\kappa_{\rm bf,0}}{\kappa_{\rm ff,0}}
\right]
\frac{\xi^2}{Y^{10}}
\nonumber \\
&\times&
\Biggr[
1
+
\sqrt{2}
\frac{\alpha_{\rm SS}}{\dot{m}}
(C_{\rm gap}a_0)^{3/2}
r_{\rm S}
\frac{\kappa_{\rm es}^2}{\kappa_{\rm ff,0}}
\left(
\frac{T_0}{1+x_e}
\right)^{7/2}
\nonumber \\
&\times&
\left[
\frac{1}{
x_e^2 + (1-x_e)
(\kappa_{\rm bf,0}/\kappa_{\rm ff,0})
}
\right]
\frac{Y^{10}}{\xi^2}
\Biggr]^{1/2} .
\label{eq:ka3}
\end{eqnarray}

Then, the optical depth for each opacity is given by
\begin{eqnarray}
\tau_1
&=&
\frac{1}{2}\kappa_1\Sigma
=
\tau_{10}\frac{1}{Y^{2}\xi^{1/2}}, 
\label{eq:tau1} \\
\tau_2
&=&
\frac{1}{2}\kappa_2\Sigma
=
\tau_{20} \frac{\xi^{1/2}}{Y^{7}},
\label{eq:tau2} \\
\tau_3
&=&
\frac{1}{2}\kappa_3\Sigma
=
\tau_{30}\frac{\xi^{3/2}}{Y^{12}}\Biggr[1+\tau_{31}\frac{Y^{10}}{\xi^2}\Biggr]^{1/2},
\label{eq:tau3}
\end{eqnarray}
respectively, where
\begin{eqnarray}
    \tau_{10}
    &=&
    \frac{1}{\sqrt{2}}\frac{\dot{m}}{\alpha_{\rm SS}}\frac{1}{(C_{\rm gap}a_0)^{1/2}},
    \nonumber \\
    \tau_{20}
    &=&
    \frac{1}{2^{3/4}}\left(\frac{\dot{m}}{\alpha_{\rm SS}}\right)^{3/2}\frac{1}{(C_{\rm gap}a_0)^{5/4}}   
    \left(\frac{\kappa_{\rm ff,0}}{r_{\rm S}}\right)^{1/2}
    \frac{x_e}{\kappa_{\rm es}} 
\left(
\frac{1+x_e}{T_0}
\right)^{7/4},
    \nonumber \\
    \tau_{30}
    &=&
    \frac{1}{2}\left(\frac{\dot{m}}{\alpha_{\rm SS}}\right)^2\frac{1}{(C_{\rm gap}a_0)^{2}}
    \frac{\kappa_{\rm ff,0}}{\kappa_{\rm es}^2}
   \frac{1}{r_{\rm S}}
\left(
\frac{1+x_e}{T_0}
\right)^{7/2}
\nonumber \\
&\times&
\left[
x_e^2 + (1-x_e)
\frac{\kappa_{\rm bf,0}}{\kappa_{\rm ff,0}}
\right],
    \nonumber \\
    \tau_{31}
    &=&
    \sqrt{2}\frac{\alpha_{\rm SS}}{\dot{m}}(C_{\rm gap}a_0)^{3/2}
    \frac{\kappa_{\rm es}^2r_{\rm S}}{\kappa_{\rm ff,0}}
    \left(
\frac{T_0}{1+x_e}
\right)^{7/2}
\nonumber \\
&\times&
    \left[
\frac{1}{
x_e^2 + (1-x_e)
(\kappa_{\rm bf,0}/\kappa_{\rm ff,0})
}
\right].
    \nonumber
\end{eqnarray}

At fixed $(\xi, Y, M)$, equations (\ref{eq:kff2})–(\ref{eq:ka2}) show that the absorption opacities increase as the binary semi-major axis $a\,(=a_0\,r_{\rm S})$ becomes larger. This happens because the dynamical temperature scale decreases as $T_{0} \propto a^{-1}$ and the disk density scales as $\rho \propto a^{-3/2}$. When these scalings are inserted into the Kramers-type Rosseland means, one finds that $\kappa_{\rm ff}$ and $\kappa_{\rm bf}$ grow approximately as $\kappa_{\rm ff}, \kappa_{\rm bf} \propto a^{2}$, apart from the ionization factors $x_e$ and $1-x_e$.

\begin{figure}[ht!]
\centering
\includegraphics[scale=0.3]{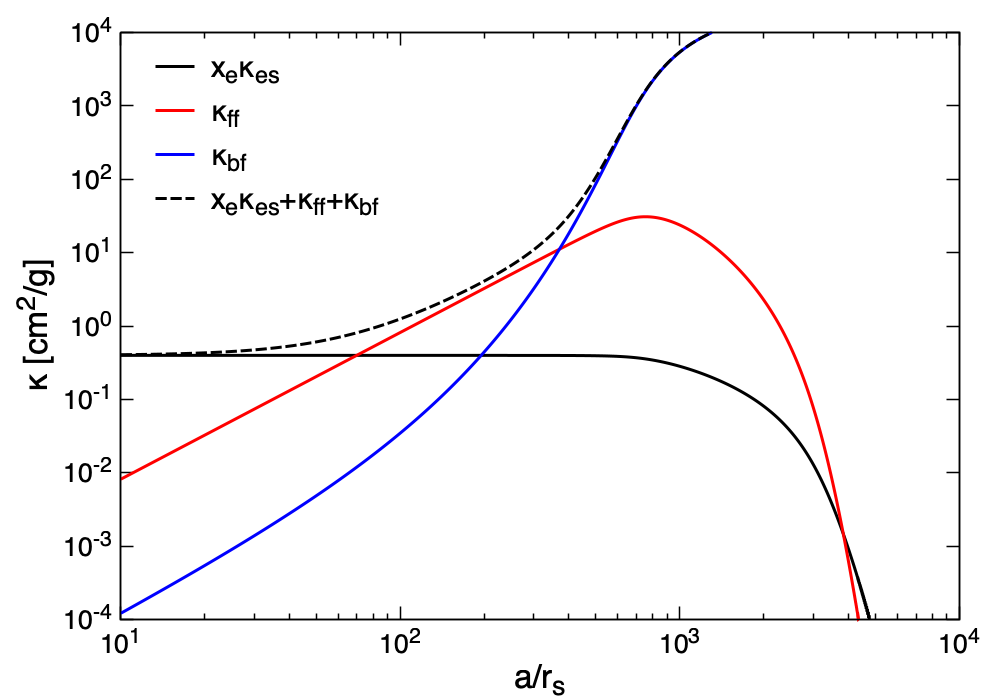}
\caption{
Analytic pure-hydrogen Rosseland-mean opacity $\kappa$ contributions as a function of the semi-major axis normalized by the Schwarzschild radius, $a/r_{\rm S}$. The opacities are evaluated for a pure-hydrogen CBD with $M=100\,M_\odot$, $Y=0.01$, and $\xi=10$. Colors denote the contributions from electron scattering multiplied by the ionization fraction ($x_e\kappa_{\rm es}$, black), free-free absorption ($\kappa_{\rm ff}$, red), and bound-free absorption ($\kappa_{\rm bf}$, blue). The black dashed curve shows the total opacity, $\kappa=x_{e}\kappa_{\rm es}+\kappa_{\rm ff}+\kappa_{\rm bf}$}.

\label{fig:opacity_vs_a}
\end{figure}

As $a/r_{\rm S}$ increases, the ionization fraction $x_e$ obtained from equation~(\ref{eq:saha}) decreases. As a result, $\kappa_{\rm ff}$ reaches a broad maximum at intermediate $a/r_{\rm S}$ and then declines, while $\kappa_{\rm bf}$ continues to rise and eventually dominates the total opacity. Figure~\ref{fig:opacity_vs_a} illustrates this behavior. At small separations, electron scattering represented by $x_{\rm e}\kappa_{\rm es}$ provides the largest contribution to the total opacity. As the binary separation increases toward $a/r_{\rm S}\sim10^{2}$, the free–free opacity $\kappa_{\rm ff}$ becomes comparable to $x_e\kappa_{\rm es}$ and both contribute significantly. For even larger separations, the bound–free opacity $\kappa_{\rm bf}$ overtakes these components and dominates $\kappa_{\rm tot}$, which then increases steeply. This is the regime relevant for our models. Thus, the dominant source of opacity in the CBD changes systematically with the binary separation.

%
\subsection{Energy equation}
\label{sec:3.2}
%

Combining equation~(\ref{eq:frad}) with equation~(\ref{eq:tct0}) yields the radiative cooling rate as a function of $\xi$.
\begin{eqnarray}
Q_{\rm rad}
=
\frac{32}{3}\frac{\sigma{T_{\rm c}^4}}{\tau}
=
\frac{32}{3}
\frac{1}{(1+x_e)^4}
\frac{\sigma{T_{\rm 0}^4}}{\tau}\frac{Y^8}{\xi^4} .
\end{eqnarray}
Also, combining equation~(\ref{eq:qvis}) with equation~(\ref{eq:dimlessparams}) yields the viscous heating rate as a function of $\xi$.
\begin{eqnarray}
Q_{\rm vis}
=
 \frac{3}{4\pi}\frac{GM\dot{M}}{r_{\rm in}^3}\frac{1}{\xi^3}.
 \label{eq:qvis2}
\end{eqnarray}

Next, the irradiation heating rate is determined by the irradiated flux on the surface of the CBD. The irradiated flux on the surface of the CBD is defined by equation~(A12) of \cite{2024ApJ...975..141L} as
\begin{equation}
F_{{\rm irr},i}
    = \frac{L_i}{4\pi d_i^2}
    (\vec l_i \cdot \vec n)(\vec n \cdot \vec e_z),
\label{eq:Firr}
\end{equation}
where $i=1$ and $i=2$ represent the primary and secondary black holes, respectively, $d_i$ is the distance between each irradiation source and the surface element on the CBD, $\vec l_i$ is the unit vector directed from the CBD surface toward the primary and secondary black holes, $\vec n$ denotes the unit vector normal to the CBD surface, and $\vec e_z$ is the unit vector along the $z$ axis (see also Appendix~A.1 of \cite{2024ApJ...975..141L} for the detailed derivation of the irradiation flux).

The azimuthally averaged irradiation heating rate is then determined by (see also equation~(16) of \cite{2024ApJ...975..141L})
\begin{eqnarray}
Q_{\rm irr}
    &=&
2\int_0^{2\pi}\,A_1F_{\rm irr,1}+A_2F_{\rm irr,2}\,d\phi
\nonumber \\
    &=&
    \frac{A_1L_1}{2\pi r_{\rm in}^2}
    \frac{1}{\xi}\Biggr[ (1+Q_{12})\frac{dY}{d\xi}-
    \frac{\beta_1+Q_{12}\beta_2}{\xi^2}
    \left(\frac{Y}{\xi} -\frac{1}{2}\frac{dY}{d\xi}
    \right)\Biggr],
    \nonumber 
    \\
\label{eq:qirr}
\end{eqnarray}
where $A_1$ and $A_2$ are the absorption fractions of the incident radiation energy from the CPD and the CSD, respectively, $Q_{12}=A_2L_2/(A_1L_1)$ with $L_1$ and $L_2$ being the bolometric luminosities defined in equation~(\ref{eq:bol_lumi}), and $\beta_1$ and $\beta_2$ are the parameters given by
\begin{eqnarray}
    \beta_1
    &=&
   \frac{3}{2}\frac{1}{C_{\rm gap}^2}\frac{q^2}{(1+q)^2},
   \\
\beta_2
&=&
      \frac{3}{2}\frac{1}{C_{\rm gap}^2}\frac{1}{(1+q)^2},
    \label{eq:beta-param}
\end{eqnarray}
respectively. Note that $\beta_1$ and $\beta_2$ are less than unity for $q\le1$. Since $A_1$ and $A_2$ are unlikely to differ significantly for two minidisks in similar physical states, we assume that $A_1=A_2=A$, resulting in $Q_{12}=q$. Moreover, we treat $A$ as a constant. In the optically thick limit, the albedo $1 - A$ is nearly independent of the local conditions and becomes constant, because multiple scattering and absorption events cause the reflected fraction to depend mainly on the ratio of scattering to absorption opacity rather than on detailed local variations \citep{2018PASJ...70...25F}. As shown in equation~(\ref{eq:qirr}), the irradiation heating rate depends not only on the radiated flux from the minidisks, represented by the term outside the brackets, but also on the flaring structure of the CBD, which appears through the terms $dY/d\xi$ and $Y/\xi$.

Therefore, with the dimensionless variables, the energy balance equation $Q_{\rm vis} + Q_{\rm irr}=Q_{\rm rad}$ is expressed as  
\begin{eqnarray}
    &&
    \frac{3}{4\pi}\frac{GM\dot{M}}{r_{\rm in}^3}\frac{1}{\xi^3}
    \nonumber \\
    &&
    +\frac{AL_1}{2\pi r_{\rm in}^2}\frac{1}{\xi}
    \left[ (1+Q_{12})\frac{dY}{d\xi}-
    \frac{\beta_1+Q_{12}\beta_2}{\xi^2}
    \left(\frac{Y}{\xi} -\frac{1}{2}\frac{dY}{d\xi}\right)\right]
    \nonumber \\
    &&
    =
    \frac{32}{3}
\frac{1}{(1+x_e)^4}
\frac{\sigma{T_{\rm 0}^4}}{\tau}\frac{Y^8}{\xi^4} .
    \label{eq:fullene}
    \end{eqnarray}

Although our equations remain 1D and axisymmetric, the binary imprint appears most clearly in the irradiation term $Q_{\rm irr}$. As defined by equation (\ref{eq:qirr}), $Q_{\rm irr}$ depends on the two minidisks’ luminosities $L_1,L_2$, the absorption factors $A_1,A_2$ (taken equal here), and the geometric coefficients $\beta_1,\beta_2$, with the ratio $Q_{12}=A_2L_2/(A_1L_1)$. The geometry of incidence, captured by the dot products in $F_{\rm{irr},\it{i}}$, together with $L_{1,2}$, $A_{1,2}$, and $\beta_{1,2}$, determines the intensity and spectral shape of the irradiation on the CBD surface. Therefore, the mass ratio $q$ and the partitioned accretion rates influence $Q_{\rm irr}$ through $L_{1,2}$, which in turn sets the outer–disk surface–temperature profile and the IR–optical SED. In our model, the CBD responds locally by satisfying the energy balance shown in equation (\ref{eq:fullene}).

%
%
\begin{table*} [htbp]
    \renewcommand{\arraystretch}{1.0}
    \centering
        \begin{tabular}{c|c|c|c|c}
            \hline
             Opacity model 
             & Opacity ($\kappa$)
             & Optical depth ($\tau$) 
             & Temperature relation 
             & Outer boundary condition ($Y_{\rm out}$) \\
            \hline
             Model~0 & $-$ & $16/3$ & $T_{\rm s}=T_{\rm c}$ & Equation (29) of \cite{2024ApJ...975..141L} \\
             Model~1 & $\kappa_{\rm es}$ & Equation~(\ref{eq:tau1}) & $T_{\rm s}\neq T_{\rm c}$ & Equation (\ref{eq:yout1}) \\
             Model~2 & $\sqrt{\kappa_{\rm es}\kappa_{\rm ff}}$ & Equation~(\ref{eq:tau2}) & $T_{\rm s}\neq T_{\rm c}$ & Equation (\ref{eq:yout2}) \\
             Model~3 & $\sqrt{\kappa_{\rm a}\left(\kappa_{\rm a}+\kappa_{\rm es}\right)}$ & Equation~(\ref{eq:tau3}) & $T_{\rm s}\neq T_{\rm c}$ & Equation (\ref{eq:yout3}) \\
            \hline
        \end{tabular}
    \caption{
    Summary of the models incorporating three different opacity prescriptions. The columns, from left to right, list the model identifier, the adopted opacity $\kappa$, the expression for the optical depth $\tau$, the relation between the surface temperature $T_{\rm s}$ and midplane temperature $T_{\rm c}$, the outer boundary condition $Y_{\rm out}$. Model~0 reproduces the fiducial case from \cite{2024ApJ...975..141L}, assuming $T_{\rm s} = T_{\rm c}$ without employing any physical opacity. Model 1 adopts pure electron scattering opacity. Model 2 describes the regime where electron scattering dominates over free-free absorption. Model 3 includes an effective opacity accounting for both free-free and bound-free absorption and electron-scattering. For all models, only gas pressure is considered; radiation and magnetic pressures are neglected.}
    \label{tbl:opmodels}
\end{table*}

%
\subsection{Basic equation}
%
Equation~(\ref{eq:fullene}) leads to the differential equation to determine the radial distribution of the disk aspect ratio as    
\begin{eqnarray}
    \frac{dY}{d\xi}
    =\left[\frac{16}{3}\frac{\alpha}{\tau}
    \frac{1}{(1+x_e)^4}
    \frac{Y^8}{\xi^3}+\beta\frac{Y}{\xi^3} -\gamma\frac{1}{\xi^2}\right]
    \left[1+\frac{\beta}{2}\frac{1}{\xi^2}\right]^{-1},
    \nonumber \\
    \label{eq:fulldiff}
\end{eqnarray}
where we introduced the following two parameters:
\begin{eqnarray}
    \alpha
    &&
    =
    \frac{1}{A}\frac{L_0}{L_1+L_2}
    ,
    \label{eq:alpha}
    \\
    \beta
    &&
    =
    \beta_1\frac{L_1}{L_1+L_2}
    +
    \beta_2\frac{L_2}{L_1+L_2}
    \label{eq:beta}
\end{eqnarray}
with the normalization luminosity defined by
\begin{eqnarray}
L_{0}
&&
\equiv 4\pi r_{\rm in}^2
    \sigma{T}_{0}^4
    \nonumber \\
    &&
    \sim 1.3\times10^{56}\,{\rm ergs^{-1}}
    \left(\frac{M}{100\,M_\odot}\right)^{2}
    \left(\frac{C_{\rm gap}}{2}\right)^{-2}
    \left(\frac{a_0}{1000}\right)^{-2}.
    \nonumber \\
    \label{eq:l0} 
\end{eqnarray}
Here, $\alpha$ is the ratio of the blackbody luminosity at the inner edge of the CBD to the total irradiation luminosity from the two minidisks, while $\beta$ is an average of $\beta_1$ and $\beta_2$, weighted by the ratio of the total irradiation to respective irradiation luminosities. Also, $\gamma$ is defined as
\begin{eqnarray}
\gamma
&&
=\frac{3}{2}\frac{1}{AL_1}
    \frac{1}{1+Q_{12}}
    \frac{GM\dot{M}}{r_{\rm in}}
    =\frac{9}{2}\frac{1}{A}
    \frac{1}{C_{\rm gap}a_0}
    \nonumber \\
    &&
\sim
2.3\times10^{-2}
\left(\frac{A}{0.1}\right)^{-1}
\left(\frac{C_{\rm gap}}{2}\right)^{-1}
\left(\frac{a_0}{1000}\right)^{-1}.
\label{eq:gamma}
\end{eqnarray}

Since equation~(\ref{eq:gamma}) indicates that $\gamma/\xi^2$ goes to zero for $\xi\gg1$, equation~(\ref{eq:fulldiff}) is reduced to 
\begin{eqnarray}
    \frac{dY}{d\xi}
    =
    \left[
    \frac{16}{3}\frac{\alpha}{\tau}
    \frac{1}{(1+x_e)^4}
    \frac{Y^8}{\xi^3}
    +
    \beta
    \frac{Y}{\xi^3} 
    \right]
    \left[
    1+\frac{\beta}{2}\frac{1}{\xi^2}
    \right]^{-1}.
\label{eq:dydx}
\end{eqnarray}
This equation is analytically solvable for a given optical depth and a fixed value of $x_e$, yielding the outer boundary conditions to solve equation~(\ref{eq:fulldiff}) numerically:
\begin{eqnarray}
Y_{\rm out}\left(\xi;\tau_{10}\right)
    &=& 
    \left(\frac{\tau_{10}}{32\alpha}\right)^{1/9}\,
    \xi_{\rm out}^{1/6}
    \left[1-\frac{11}{42}\frac{\beta}{\xi_{\rm out}^2}\right],
    \label{eq:yout1} 
    \\
Y_{\rm out}\left(\xi;\tau_{20}\right)
    &=& 
    \left(\frac{15}{448}\frac{\tau_{20}}{\alpha}\right)^{1/14}\,
    \xi_{\rm out}^{5/28}
    \left[1-\frac{17}{84}\frac{\beta}{\xi_{\rm out}^2}\right],
    \label{eq:yout2}
    \\
Y_{\rm out}\left(\xi;\tau_{30}\right)
    &=& 
    \left(\frac{21}{608}\frac{\tau_{30}}{\alpha}\right)^{1/19}\,
    \xi_{\rm out}^{7/38}
    \left[1-\frac{69}{418}\frac{\beta}{\xi_{\rm out}^2}\right],
    \nonumber 
    \\
    \label{eq:yout3}
\end{eqnarray}
where $\xi_{\rm out}=r_{\rm out}/r_{\rm in}$ with $r_{\rm out}$ being the outer radius of the CBD. For details, see Appendix A, where approximate solutions for $\xi \gg 1$ are given by equations~(\ref{eq:anaytau1})-(\ref{eq:anaytau3}). The explicit form of $\tau_{10}, \tau_{20}, \tau_{30}$ are given by equations~(\ref{eq:tau10})-(\ref{eq:tau30}), respectively. The radial temperature profile is then obtained by numerically solving equation~(\ref{eq:tct0}) with the numerical solutions for equation~(\ref{eq:fulldiff}).

%
\subsection{Parameter dependence of CBD spectral shape}
\label{sec:paramsdep}
%

Our model has eight parameters:
$q$, $\alpha_{\rm SS}$, $A$, $C_{\rm gap}$, $\dot{m}$, $M$, $a_0$, and $\xi_{\rm out}$. Throughout this paper, the fiducial values of these parameters are set as follows: $q=1$, $\alpha_{\rm SS}=0.1$, $C_{\rm gap}=2$, $A=0.1$, $\dot{m}=1$, $M=100\,M_\odot$, $a_0=1000$, and $\xi_{\rm out}=10^4$.

For later use it is convenient to summarize how the heating rates depend on these model parameters. Near the inner edge, where viscous heating dominates, the vertically integrated viscous flux scales as
\begin{equation}
  Q_{\rm vis} \propto 
  \dot{m}\, M^{-1}\, (C_{\rm gap}\,a_{0})^{-3},
  \label{eq:qvis_app}
\end{equation}
where equation~(\ref{eq:qvis2}) is used. In the irradiation–dominated regime with the $\tau_{3}$ opacity, $Q_{\rm irr}\propto T_{\rm s}^{4}$ with 
the asymptotic solution of the outer boundary condition gives
\begin{eqnarray}
  Q_{\rm irr} 
  &\propto&
\alpha_{\rm SS}^{-2/19}\,
  A^{20/19}\,
  \dot{m}^{22/19}\,
  \nonumber 
  \\
  &\times&
  M^{-21/19}\,
  (C_{\rm gap} a_{0})^{-69/38}\,
\xi_{\rm out}^{-69/38},
  \label{eq:qirr_app}
\end{eqnarray}
where $x_e=1$ is adopted for simplicity, and equations~(\ref{eq:frad}), (\ref{eq:tct0}), and (\ref{eq:yout3}) were used for the derivation. The relative importance of each parameter for the spectral shape is essentially determined by the exponents in the above two scalings.

The strongest dependences in equation~(\ref{eq:qirr_app}) arise from the geometric and global scalings, in particular $(C_{\rm gap}a_0)^{-69/38}$ and subsequently $M^{-21/19}$. Equation~(\ref{eq:qirr_app}) also shows that $Q_{\rm irr}(\xi_{\rm out}) \propto \xi_{\rm out}^{-69/38}$ for the local irradiation heating rate evaluated near the outer boundary when the other parameters are fixed. This negative exponent reflects that the irradiation flux intercepted per unit area decreases toward larger radii. Nevertheless, $\xi_{\rm out}$ remains important for the observable SED because it sets the maximum emitting radius and therefore the amount of cool reprocessing area included in the radial integration. Increasing $\xi_{\rm out}$ adds cooler annuli that dominate the lowest-frequency emission and broadens the IR bump toward lower frequencies.

The geometric factor $C_{\rm gap} a_{0}$ enters the scalings of both heating terms with a large negative exponent. Because of this scaling, larger values of $C_{\rm gap} a_{0}$ reduce both the viscous heating flux $Q_{\rm vis}$ and the irradiative heating flux $Q_{\rm irr}$ at a given dimensionless radius. In practice, however, $C_{\rm gap}$ is expected to vary only within a narrow range of order unity, whereas the binary separation $a_{0}$ can change by orders of magnitude. For this reason, the variation of $a_{0}$ is much more important than that of $C_{\rm gap}$. We therefore fix $C_{\rm gap}=2$ as a representative value (see equation~(\ref{eq:rin})) and explicitly compare CBD spectra corresponding to $a_{0}=100$ and $a_{0}=1000$ for our four opacity models.

The binary mass $M$ enters the above scalings of both $Q_{\rm vis}$ and $Q_{\rm irr}$ with a moderate exponent. More importantly, $M$ sets $r_{\rm in}$ and determines the global frequency scale of the CBD emission. We therefore explore the dependence on $M$ in Section~\ref{sec:cbdspectra}, while keeping the other parameters fixed, in order to separate the effect of mass scaling from the opacity-driven changes in the spectral shape.

The parameters $\dot{m}$ and $A$ enter $Q_{\rm irr}$ with moderately large positive exponents, whereas only $\dot{m}$ enters $Q_{\rm vis}$ and it does so linearly. They mostly rescale the overall luminosity of the irradiated disk and only mildly modify the spectral shape in the outer, irradiation dominated region. Similarly, the Shakura–Sunyaev parameter $\alpha_{\rm SS}$ appears in $Q_{\rm irr}$ with a very small exponent and does not enter $Q_{\rm vis}$ explicitly in our formulation, so it has a negligible effect on the spectral shape in this regime. Therefore, for the purpose of illustrating the qualitative differences between the opacity models, we fix $\dot{m}=1$, $A=0.1$, and $\alpha_{\rm SS}=0.1$ in the present work.

Finally, the mass ratio $q$ appears only through $\beta$ and produces only a higher order correction to the irradiation flux. Its influence on the spectral shape of the outer disk is much weaker than that of $\xi_{\rm out}$ or $a_{0}$, and it does not affect $Q_{\rm vis}$ in our prescription. This very weak dependence is consistent with Figure 3(b) of \cite{2024ApJ...975..141L}, where the CBD spectra for different mass ratios almost overlap.

Given these scalings, we fix $(q, \alpha_{\rm SS}, A, \dot{m}, C_{\rm gap})$ to their fiducial values in the rest of this paper. We treat $\xi_{\rm out}$, $M$, and $a_0$ as the key parameters that control the CBD spectral shape. In Section~\ref{sec:cbdspectra} we present CBD spectra that illustrate the dependence on $M$ and $\xi_{\rm out}$, and we also compare spectra for $a_0 = 1000$ and $a_0 = 100$ to quantify the effect of the binary semi-major axis.

%
\section{Results}
\label{sec:results}
%

In this section, we provide the numerical solutions for the radial temperature distribution of the CBD and the corresponding spectrum, and discuss how they depend on the opacity models. We also calculate the triple disk spectrum and compare it with the observational detection limits.

%
\subsection{Circumbinary Disk Temperature Profiles}
\label{sec:cbdtemp}
%

The midplane temperature, $T_{\mathrm c}(r)$, is obtained by numerically integrating equation~(\ref{eq:fulldiff}) for the CBD aspect ratio, inserting the solution into equation~(\ref{eq:tct0}), and solving it. The corresponding surface temperature, $T_{\mathrm s}(r)$, is then obtained from the radiative-equilibrium relation given by equation~(\ref{eq:tstc}). We evaluate four opacity prescriptions (Models~0–3; see Table~\ref{tbl:opmodels}) in order to isolate the contribution of electron-scattering, free–free, and bound–free absorption to the thermal structure of CBD.

\begin{figure*}[htbp]
    \centering
\includegraphics[scale=0.25]{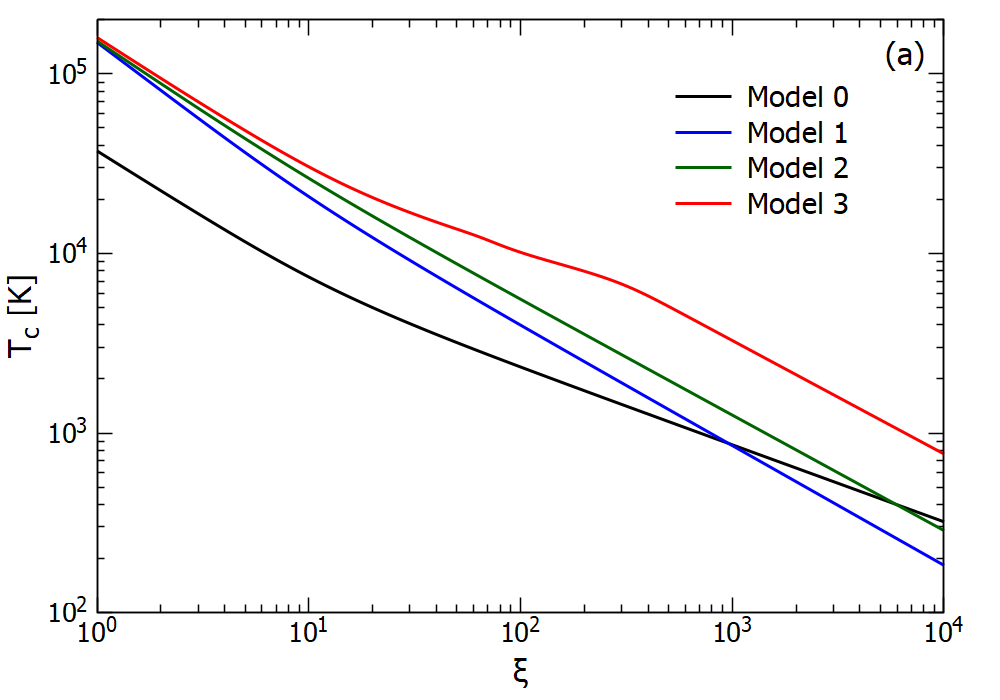}
\includegraphics[scale=0.25]{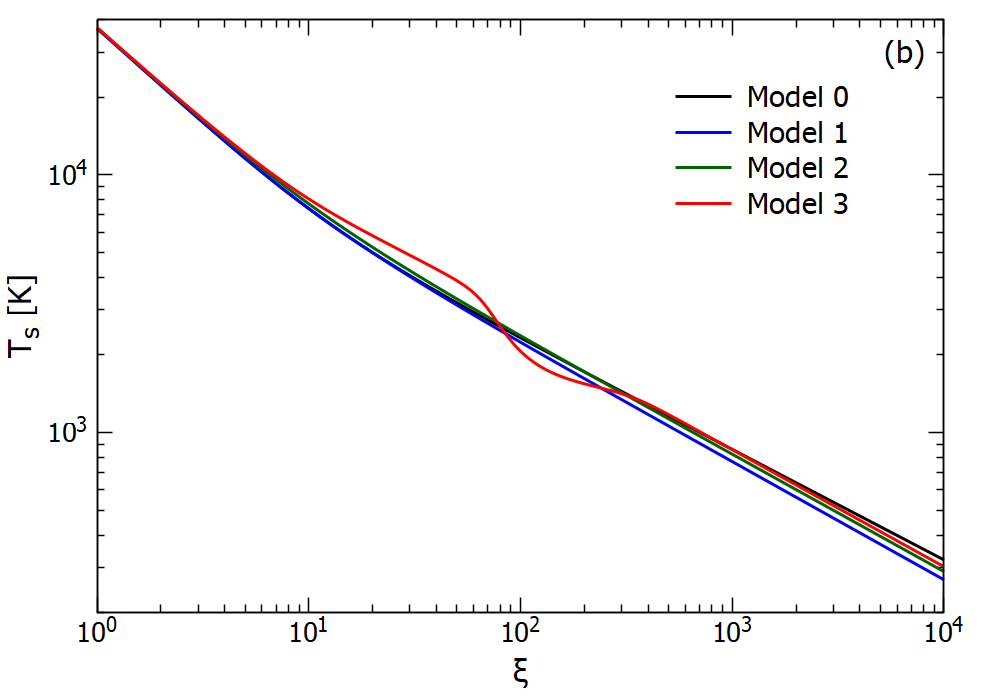}
    \caption{
    Radial temperature profiles of the CBD for the four opacity models in Table~\ref{tbl:opmodels}. Panels (a) and (b) show the midplane temperature $T_{\mathrm c}$ and the surface temperature $T_{\mathrm s}$, respectively. Both quantities are plotted against the normalized radius $\xi$ on logarithmic axes. Solid curves correspond to Model 0 (black), Model 1 (blue), Model 2 (green), and Model 3 (red).
        }
   \label{fig:temp}
\end{figure*}

Figure~\ref{fig:temp} compares $T_{\rm c}$ and $T_{\rm s}$ for the four models. Panel (a) shows that due to the opacity effect, the midplane temperatures $T_\mathrm{c}$ of Models 1-3 is significantly higher than that of Model 0, except in the outermost regions for Models 1 and 2. Among Models 1-3, the difference of $T_\mathrm{c}$ increases with $\xi$, as both the free-free and bound-free absorptions increase with decrease of disk temperature. In contrast, panel (b) demonstrates that the surface temperatures are almost indistinguishable for $\xi\lesssim10$, where viscous heating dominates and is effectively independent of opacity. Irradiation affects only layers near the surface; hence subtle, model-dependent differences appear at larger radii, where irradiation becomes comparable to, or exceeds, viscous heating.

Disk opacity alters the midplane temperature profile more substantially than the surface temperature profile. Since the emergent spectrum is determined by $T_{\rm s}$, only modest spectral differences between the models are expected, primarily in the outer regions where irradiation heating dominates. The subsequent section quantifies these differences and explores their observational consequences.

%
\subsection{Circumbinary Disk Spectra}
\label{sec:cbdspectra}
%

Because the CBD is optically thick in the vertical direction, its surface emits local blackbody radiation with spectral intensity given by
\begin{eqnarray}
    I_\nu=\frac{2h}{c^2}\frac{\nu^3}{\exp(h\nu/kT_{\rm s})-1},
    \nonumber 
\end{eqnarray}
where $h$ is the Planck constant, $k$ is the Boltzmann constant, and $\nu$ is the frequency. The flux density to be emitted from the whole CBD surface is then given by \citep{kato_black-hole_2008}
\begin{eqnarray}
S_{\nu}
=
\int I_\nu\,
d\Omega
=
4\pi
\frac{h}{c^2}
\frac{\cos{\delta}}{D_{\rm L}^2}
\nu^3
\int_{r_{\rm in}}^{r_{\rm out}}
\,
\frac{r}{e^{h\nu/(kT_{\rm s})}-1}
\,dr,
\nonumber \\
\label{eq:snu}
\end{eqnarray}
where $\delta$ is the inclination angle of the disk and $D_{\rm L}$ is the luminosity distance. In the following we adopt $\delta=0$ unless otherwise noted.

%
%
\begin{figure*}[http]
    \centering
\includegraphics[scale=0.25]{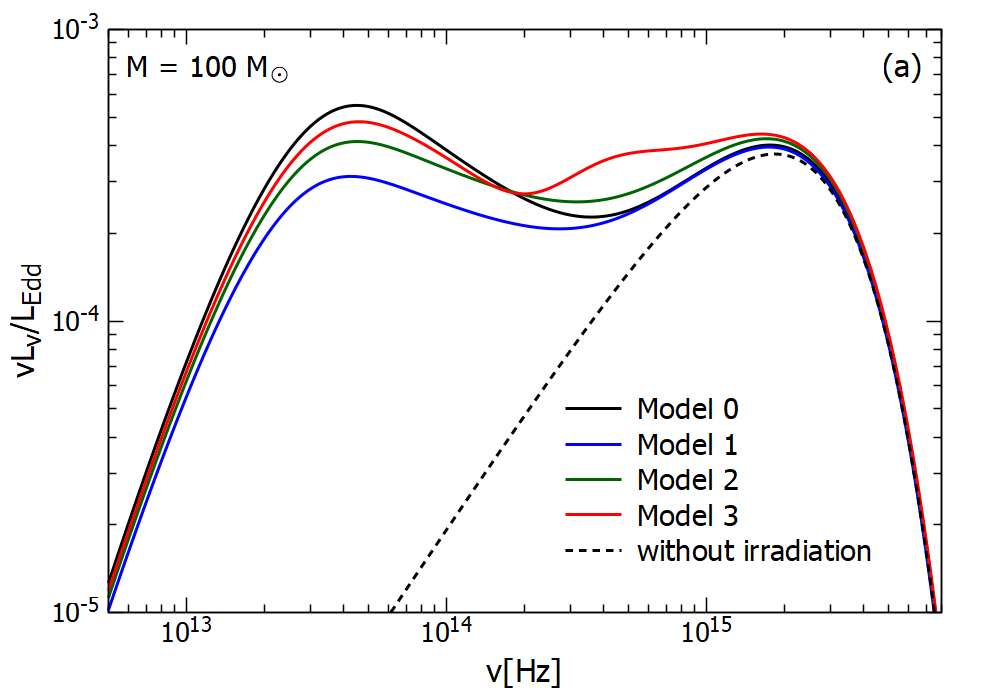}
\includegraphics[scale=0.25]{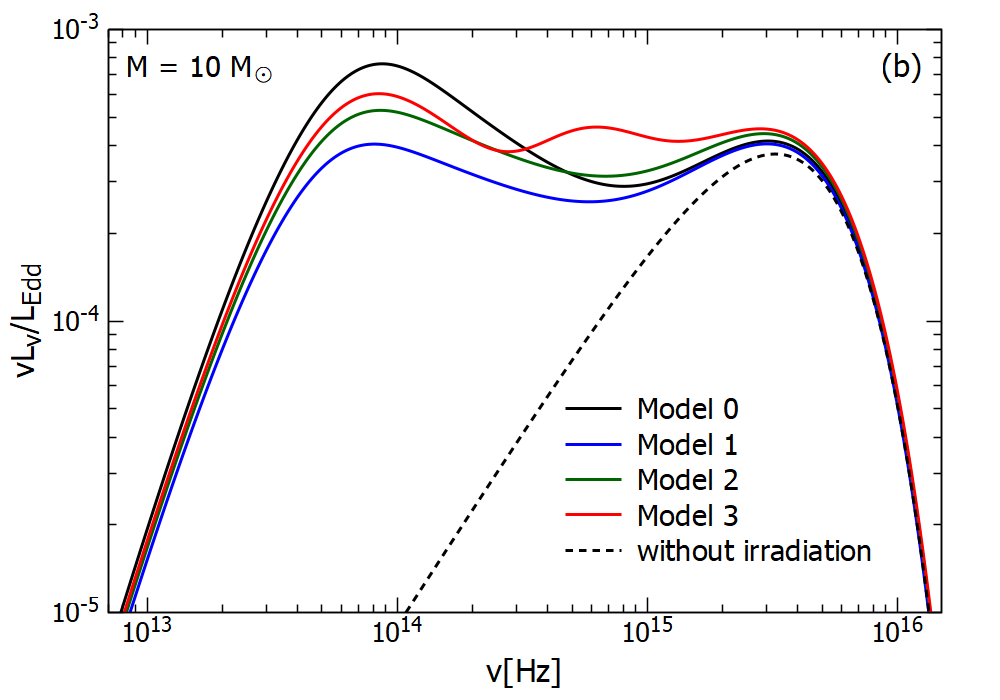}
\includegraphics[scale=0.25]{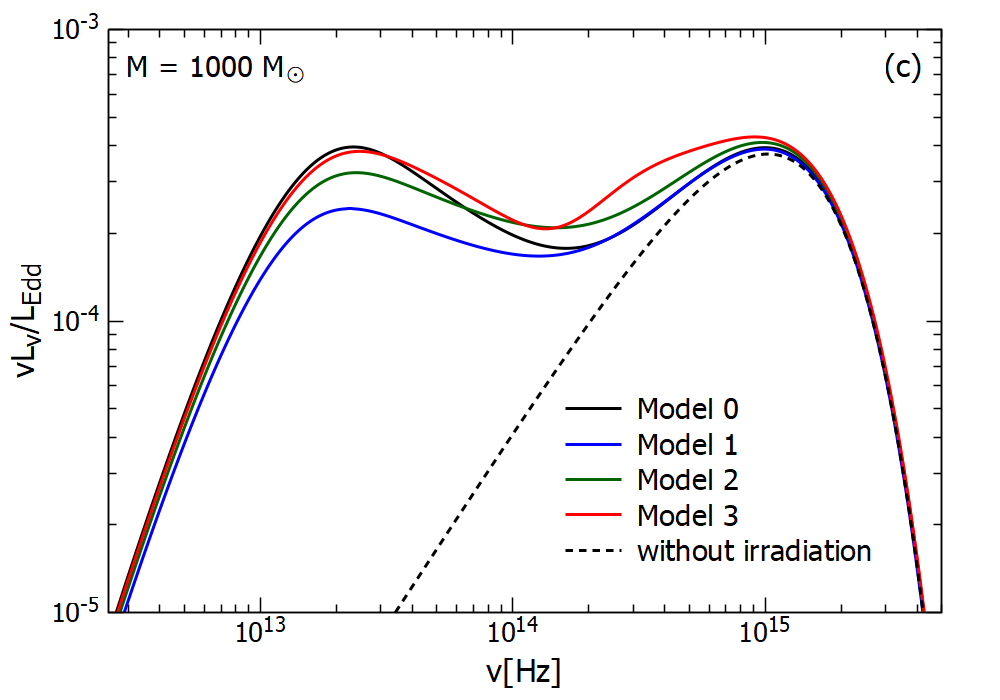}
    \caption{
SEDs of the CBD for the four opacity models listed in Table~\ref{tbl:opmodels}. Panels (a)–(c) correspond to total binary masses of $100\,M_\odot$ (fiducial case), $10\,M_\odot$, and $1000\,M_\odot$, respectively. Each panel plots the normalized luminosity $\nu L_\nu/L_{\rm Edd}$ versus frequency on logarithmic axes. Solid curves indicate Model~0 (black), Model~1 (blue), Model~2 (green), and Model~3 (red); the dashed black curve reproduces Model~0 without external irradiation. 
    }
   \label{fig:CBD_mass}
\end{figure*}

\begin{figure*}[http]
    \centering
\includegraphics[scale=0.25]{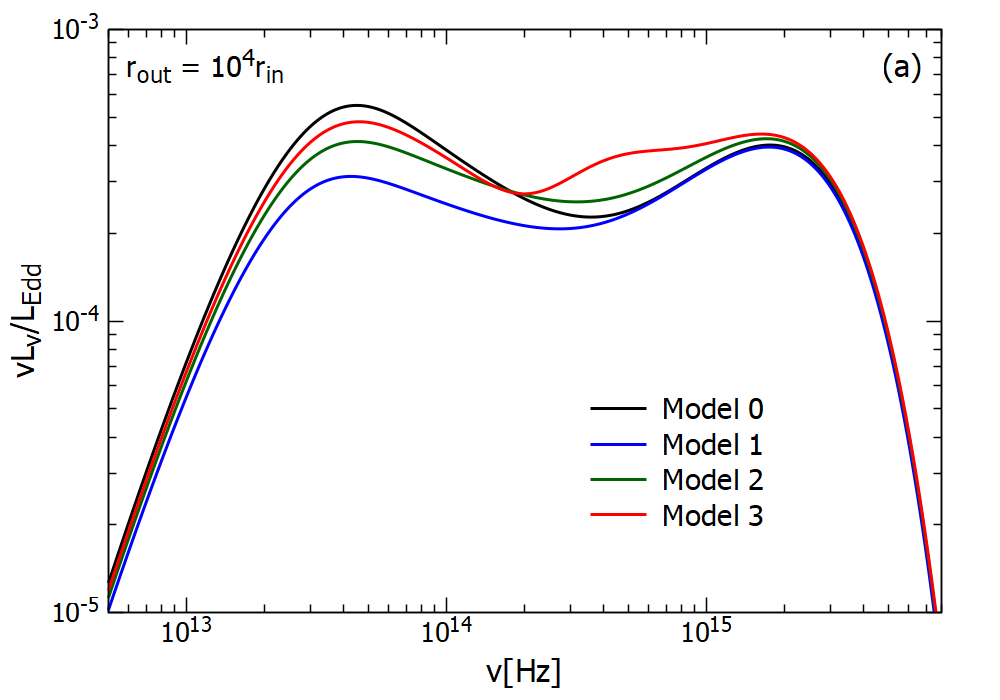}
\includegraphics[scale=0.25]{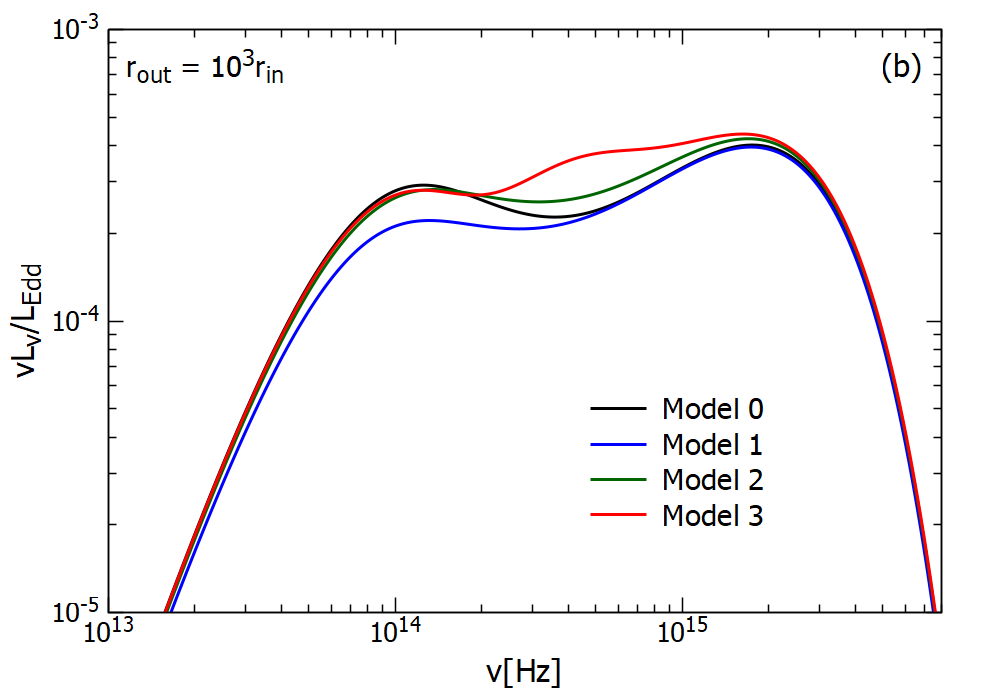}
\includegraphics[scale=0.25]{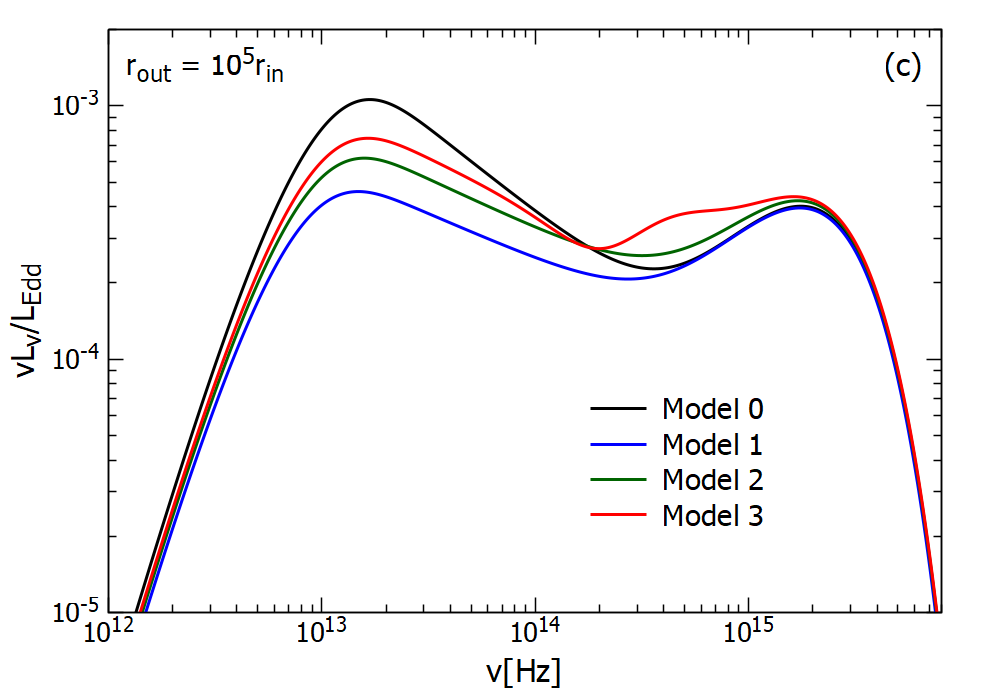}
    \caption{
Same format as Figure~\ref{fig:CBD_mass} but for different outer disk radii, varying the outer radius $\xi_{\rm out}$ while keeping the binary mass fixed at $M=100\,M_\odot$. Panels (a)–(c) adopt $\xi_{\rm out}=10^{4}$ (fiducial case), $10^{3}$, and $10^{5}$, respectively; all other parameters are identical to those in Figure~\ref{fig:CBD_mass}.
    }
   \label{fig:CBD_rout}
\end{figure*}

In what follows, we only vary the model parameters that most strongly affect the CBD spectral shape, namely the binary mass $M$, the dimensionless outer radius $\xi_{\rm out}$, and the dimensionless semi-major axis $a_{0}$, as identified by the scalings of the viscous and irradiation heating rates in Section~\ref{sec:paramsdep} (equations~(\ref{eq:qvis_app}) and (\ref{eq:qirr_app})). The remaining parameters $(q, \alpha_{\rm SS}, A, \dot{m}, C_{\rm gap})$ are kept fixed at their fiducial values, and we examine how the CBD spectra respond to changes in $M$, $\xi_{\rm out}$, and $a_{0}$ for the four opacity prescriptions summarized in Table~\ref{tbl:opmodels}.

Figure \ref{fig:CBD_mass} illustrates how the emergent CBD spectrum responds to both the total binary mass and the choice of opacity prescription. For our fiducial case of $M=100\,M_\odot$ (Panel~a), the spectrum shows the double-peaked shape, with the low-frequency peak produced by reprocessed irradiation in the outer disk and the high-frequency peak set by viscous heating in the inner regions. Introducing bound-free and free-free absorption (Model 3, red) flattens the IR-optical continuum; the height of the low-frequency peak remains within $\lesssim0.1$ dex of the opacity-free spectrum (Model 0, black); Models 1 and 2 occupy intermediate positions and their spectral shapes are closer to that of Model 0, confirming that free-free absorption is the principal driver of this deviation. 

The dashed black line represents the spectrum of the CBD without irradiation heating. In this case, the low-frequency peak is entirely absent, and the spectrum is dominated by the inner disk's thermal emission powered purely by viscous dissipation. This remarkable difference underscores the role of irradiation in shaping the outer disk emission; without it, the surface temperature in the outer regions remains too low to contribute significantly to the IR-optical band. The contrast between the dashed and solid curves highlights that the low-frequency enhancement in the irradiated models arises from externally reprocessed radiation, rather than from local viscous heating.

Varying the binary mass, $M$, primarily shifts the entire SED along the frequency axis. For a lower-mass system ($M = 10\,M_\odot$; panel~b), both the irradiation-induced IR peak and the viscously heated UV shoulder shift to higher frequencies. In contrast, a higher-mass system ($M = 1000\,M_\odot$; panel~c) produces a cooler and more extended CBD, moving both features to lower frequencies. These mass-dependent shifts are consistent with the scaling relations derived in Section~\ref{sec:paramsdep}. Because the CBD inner radius scales linearly with $M$ (see equation~(\ref{eq:rin})), the surface temperatures in both the viscously heated inner region and the irradiation-dominated outer region follow the trends implied by equations~(\ref{eq:qvis_app}) and (\ref{eq:qirr_app}). In summary, a smaller binary mass gives a more compact and hotter disk whose spectral features move to higher frequencies, whereas a larger binary mass gives a cooler and more extended CBD whose emission peaks at lower frequencies. Thus, the binary mass mainly sets the overall frequency scale of the CBD emission, while the basic spectral shape remains almost unchanged.

At high frequencies ($\nu \gtrsim 10^{15}\,\mathrm{Hz}$), all four models yield nearly identical spectra. In this regime, viscous heating dominates, and the surface temperature is governed by the local energy dissipation rate, making the spectra largely independent of the assumed opacity. The tiny residual offsets in the high-frequency peak arise because frequencies near the peak still receive contributions from radii where irradiation and disk flaring are comparable to viscous heating. At these transition radii, the surface temperatures depend weakly on the opacity model and on the outer boundary condition, shifting the integrated high-frequency luminosity very slightly. At lower frequencies, by contrast, the IR peak does not change monotonically from one model to another. Model~0, which assumes $T_{\mathrm s} = T_{\mathrm c}$ and neglects all physical opacity sources, gives the brightest IR emission because it sets the surface temperature equal to the hotter midplane temperature and does not include any absorption that would reduce the escaping flux. Among the models that include opacity, Model~3 has the strongest IR flux, followed by Model~2 and then Model~1. This ordering reflects differences in the surface temperature profiles (see Figure~\ref{fig:temp}), which are set by how efficiently each model absorbs and thermalizes the irradiation in the outer part of the CBD.

In Model~3, the inclusion of bound-free absorption greatly increases the Rosseland mean opacity in the temperature range $T \sim 10^3$--$10^4\,\mathrm{K}$, which coincides with the irradiated outer CBD regions where the disk flares (i.e., the aspect ratio increases with radius). This enhances the local trapping of irradiating photons and raises the surface temperature, leading to stronger IR re-emission than in Model~2, where free-free absorption is included together with electron scattering but bound-free absorption is absent. In contrast, electron scattering only in Model~1 does not effectively trap irradiating photons due to the weaker disk flaring, and the surface therefore remains cooler with a lower IR flux. Taken together, these results indicate that observational constraints on CBD opacity sources are most effective in the IR for high-mass binaries and in the optical for low-mass binaries, where the spectral differences among models are most pronounced.

In panel~(b) of Figure~\ref{fig:CBD_mass}, the spectrum of Model~3 (red curve) clearly exceeds those of the other three models at frequencies $\nu \gtrsim 10^{14.4}\mathrm{Hz}$, spanning the optical and near-UV bands. At lower frequencies ($\nu \lesssim 10^{14.4}\mathrm{Hz}$), Model~0 (black curve), which includes no physical absorption opacity, is the brightest model among the four. Within this panel, the spectra therefore show a clear two-level pattern: Model~0 is brighter at lower frequencies, whereas Model~3 dominates at higher frequencies. Furthermore, when the SEDs of Model~3 are compared across the three mass cases (panels (a), (b), and (c)), the $10M_\odot$ case exhibits the highest normalized luminosity.

This behavior is interpreted as follows. In the $10M_\odot$ case, the outer surface layers of the disk cool to temperatures below $10^4\,\mathrm{K}$, where hydrogen is partially ionized and bound–free opacity is especially strong. High-energy photons emitted by the two minidisks are absorbed in these outer layers, thermalized, and re-emitted at IR wavelengths, which produces a pronounced IR peak. At the same time, the inner disk remains hot enough that its surface temperature does not decrease significantly, and the mid- and high-frequency emission is enhanced. As a result, the SED of Model~3 develops an apparently “triple-peaked” profile, although this shape is best understood as a modification of the basic double-peaked SED structure. The two main peaks are produced by viscous heating in the inner CBD and by reprocessed emission from the irradiated outer disk, and bound–free absorption in Model~3 adds a smaller subpeak in the mid-frequency range by distorting the irradiated component of the spectrum and creating a visible peak between the main IR peak and the UV shoulder. It is therefore most appropriate to regard the SED as fundamentally double-peaked, with an additional mid-frequency peak caused by bound–free opacity effects.

Figure~\ref{fig:CBD_rout} shows how the CBD spectrum changes when the outer disk radius $r_{\rm out}$ varies. In this figure, panels~(a), (b), and (c) correspond to $r_{\rm out} = 10^4\,r_{\rm in}$ (fiducial case), $10^3\,r_{\rm in}$, and $10^5r_{\rm in}$, respectively. The black, blue, green, and red curves represent Models~0, 1, 2, and 3, respectively. As $r_{\rm out}$ increases, the irradiated area of the outer CBD becomes larger, the reprocessed radiation from the two minidisks is absorbed and re-emitted over a wider region, and the IR peak becomes stronger; panel~(c), which has the largest outer radius, therefore shows the most prominent IR peak, whereas panel~(a), with the smallest radius, shows the weakest IR excess. The spectral differences between the models also become more noticeable in the IR region as $r_{\rm out}$ increases, because the opacities adopted in each model mainly control how efficiently the irradiating photons are reprocessed. When irradiation covers a larger fraction of the disk, these opacity effects are amplified and lead to clearer separations among the model spectra.

These trends are also consistent with the scaling of the irradiation heating rate derived in Section~\ref{sec:paramsdep}. In the $\tau_3$-opacity regime that we use to describe the outer CBD, equation~(\ref{eq:qirr_app}) shows that the local irradiation heating rate evaluated at the outer boundary decreases as $Q_{\rm irr}(\xi_{\rm out}) \propto \xi_{\rm out}^{-69/38}$ when the other parameters are kept fixed. Since a larger outer radius increases the reprocessing area and includes cooler annuli that dominate the lowest-frequency emission, $\xi_{\rm out}$ mainly sets how far the IR bump extends to low frequencies and, through the radial integration, how strong it appears the overall irradiation-powered IR bump.

Finally, we examine how the CBD spectrum depends on the binary semi-major axis at a fixed physical outer radius. As equation (\ref{eq:rin}) shows, the inner edge of the CBD scales with the semi-major axis as $r_{\rm in} = C_{\rm gap} a = C_{\rm gap} a_0 r_{\rm S}$, so that the outer radius is $r_{\rm out} = \xi_{\rm out} r_{\rm in} \propto \xi_{\rm out} a_0$ for fixed $C_{\rm gap}$ and $M$. Using the fiducial model with $a_0 = 1000$ and $\xi_{\rm out} = 10^4$ as a reference, we therefore adopt $\xi_{\rm out} = 10^5$ for a more compact binary with $a_0 = 100$, so that $r_{\rm out}$ is identical in both cases. This choice allows us to isolate the effect of changing $a_0$ without introducing any change in the emitting area of the outer disk.

The dynamical coupling between the CBD and the binary is governed by whether the viscous inflow speed at the cavity rim can keep up with the GW-driven inspiral rate (see \citealt{armitage_accretion_2002,2023ApJ...949L..30D}). To verify that the binary separations explored here correspond to dynamically disk-coupled configurations, we estimate the decoupling radius by equating the GW-driven inspiral speed to the viscous inflow speed at $r_{\rm in}$ as
\begin{eqnarray}
a_{\rm dec}
&=&
\frac{1}{2}\left[
\frac{128}{15}\,
\frac{q}{(1+q)^2}\,
\frac{\sqrt{C_{\rm gap}}}{\alpha_{\rm SS}\,Y_{\rm in}^2}
\right]^{2/5}
r_{\rm S}
\nonumber \\
&\sim&
70\,r_{\rm S}\,
\left(\frac{\mathcal{Q}}{1/4}\right)^{2/5}
\left(\frac{C_{\rm gap}}{2}\right)^{1/5}
\left(\frac{\alpha_{\rm SS}}{0.1}\right)^{-2/5}
\left(\frac{Y_{\rm in}}{0.01}\right)^{-4/5}
\nonumber 
\\
\label{eq:adec}
\end{eqnarray}
(e.g., equation~(\ref{eq:rin}) of \citealt{Dorazio2025}), where $Y_{\rm in}=(H/r)_{r=r_{\rm in}}$ and $\mathcal{Q}\equiv q/(1+q)^2$. For our fiducial parameters, both $a_0=100$ and $a_0=1000$ exceed $a_{\rm dec}/r_{\rm S}$ (with $a_0=100$ being closer to the transition). Therefore, the binaries considered in this paper remain dynamically coupled to the CBD.

\begin{figure}[http]
    \centering
\includegraphics[scale=0.3]{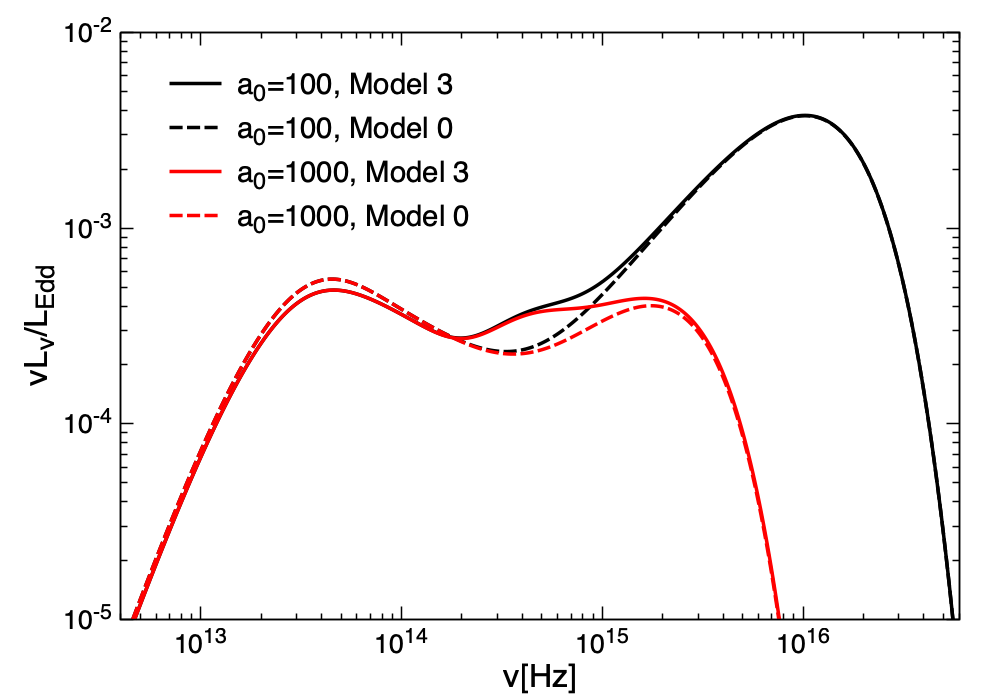}
    \caption{ 
CBD spectra for different binary separations at a fixed physical outer radius. The normalized luminosity $\nu L_\nu / L_{\rm Edd}$ is plotted as a function of frequency $\nu$. The four curves show Models 0 and 3, which are described in Table 1, for two combinations of the dimensionless parameters: $(a_0, \xi_{\rm out}) = (1000, 10^4)$ and $(100, 10^5)$. The values of $\xi_{\rm out}$ are chosen so that the physical outer radius $r_{\rm out} = \xi_{\rm out} r_{\rm in} \propto \xi_{\rm out} a_0$ is the same in both cases.
}
   \label{fig:CBD_a0}
\end{figure}

Figure \ref{fig:CBD_a0} compares the CBD spectra for Models 0 and 3, while Models 1 and 2 always lie between these two cases and are omitted for clarity. For both opacity prescriptions, decreasing $a_0$ moves the high-frequency peak to a higher frequency and increases its amplitude because the inner edge ($r_{\rm in}$) becomes smaller and hotter. This behavior is consistent with equation (\ref{eq:qvis_app}), in which the viscous heating rate $Q_{\rm vis}$ near $r_{\rm in}$ scales as $(C_{\rm gap}a_0)^{-3}$.

As noted above, we choose $(a_0,\xi_{\rm out})=(1000, 10^{4})$ and $(100, 10^{5})$ so that the physical outer radius $r_{\rm out}=\xi_{\rm out}r_{\rm in}$ is the same in both cases. Consequently, the irradiated area of the outer CBD and the characteristic incident flux $(L_{1}+L_{2})/(4\pi r_{\rm out}^{2})$ are similar for the two models, as is the total energy available for reprocessing in the CBD.

Equation (\ref{eq:qirr_app}) describes how the local irradiation heating rate $Q_{\rm irr}$ scales at large radii when $\xi_{\rm out}$ is varied at fixed model parameters. In our setup, however, $a_0$ and $\xi_{\rm out}$ are varied together so that $r_{\rm out}$ stays fixed, cancelling much of the formal dependence of $Q_{\rm irr}$ on $a_0$ and $\xi_{\rm out}$. The surface temperature responds only as $T_{\rm eff}\propto Q_{\rm irr}^{1/4}$, which further weakens the remaining dependence. In addition, the IR bump is produced by emission integrated over a broad range of radii rather than only at $r_{\rm out}$. This means that enhancements of $Q_{\rm irr}$ near the outer edge are diluted in the total spectrum. For these reasons, the height of the irradiation-powered IR bump varies only weakly between the two values of $a_0$. The small remaining differences in the IR range appear mainly in Model 3, where bound-free opacity slightly increases the reprocessing efficiency in the cool outer layers, producing a somewhat broader low-frequency shoulder. In this comparison, changing $a_0$ primarily affects the inner, viscously heated component at fixed $r_{\rm out}$. The structure of the IR bump is governed by the disk size and the opacity model rather than by $a_0$ itself.

%
\subsection{Triple disk spectra}
\label{sec:tripledisk}
%

We compute the SED of a triple disk system consisting of the CPD and CSD (hereafter ``minidisks”), and the surrounding CBD. The minidisks' SED are modeled as surface-emission integrals over local blackbodies. In our framework, the emergent multi-color blackbody spectrum from each minidisk is controlled by the surface temperature profile of each minidisk, which is determined to be proportional to $r^{-3/4}$ for given black hole mass and mass accretion rate by the local dissipation rate of the standard disk model \citep{1973A&A....24..337S,1981ARA&A..19..137P}. The pressure (gas- vs. radiation-pressure dominated) and opacity regimes modify primarily the vertical structure and midplane temperature but do not alter the resulting multi-color blackbody spectrum (e.g., \citealt{kato_black-hole_2008}).

The spectral luminosity of each minidisk is calculated using the relation $L_{\nu,i} = 4\pi D_{\rm L}^2 S_{\nu,i}$, where $S_{\nu,i}$ is the flux density emitted from the CPD ($i=1$) or the CSD ($i=2$). This flux density is computed using the same integrand as in equation (\ref{eq:snu}), but it is integrated over a distinct radial range, from the ISCO radius, $r_{{\rm in},i} = 6GM_i/c^2$, to the outer edge of each disk. The outer edge radius is defined as a fixed fraction (typically $\sim0.5$) of the Hill radius, which represents the tidal truncation limit for an accretion disk around a primary object in a binary system (e.g., \citealt{2011MNRAS.413.1447M,2021ApJ...910L..26P}). For a BBH with minidisks, their tidal truncation radii are numerically calibrated to be $r_{\rm out, \it i}\simeq 0.3-0.4\,a$ (e.g., \citealt{2014ApJ...783..134F, Miranda2017,DuffellEtAl2020,MunozEtAl2020,DOrazioDuffell2021}). At the minidisk photosphere, we approximate the opacity as electron-scattering dominated, which is appropriate for the hot minidisk temperatures ($\gtrsim 10^{5}\,{\rm K}$) in our parameter range. Electron scattering can harden the emergent spectrum relative to a blackbody \citep{ShimuraTakahara1995}. For simplicity, we neglect this effect and model each minidisk as an uncorrected multi-color blackbody.

%
%

The triple disk spectrum is then given by the sum of CBD, CPD, and CSD specific luminosities as
\begin{equation}
L_{\nu}
=
L_{\nu,{\rm CBD}}
+
L_{\nu,1}
+
L_{\nu,2}
\nonumber
\end{equation}
where $L_{\nu,{\rm CBD}}$ is obtained from the CBD vertical solution including irradiation and our opacity prescriptions described in Sections~\ref{sec:cbdtemp} and \ref{sec:cbdspectra}. In practice, the composite SED shows a high–frequency peak from the hot inner minidisks and an IR–optical excess shaped by CBD irradiation/reprocessing.

%
%
\begin{figure*}[http]
    \centering
\includegraphics[scale=0.3]{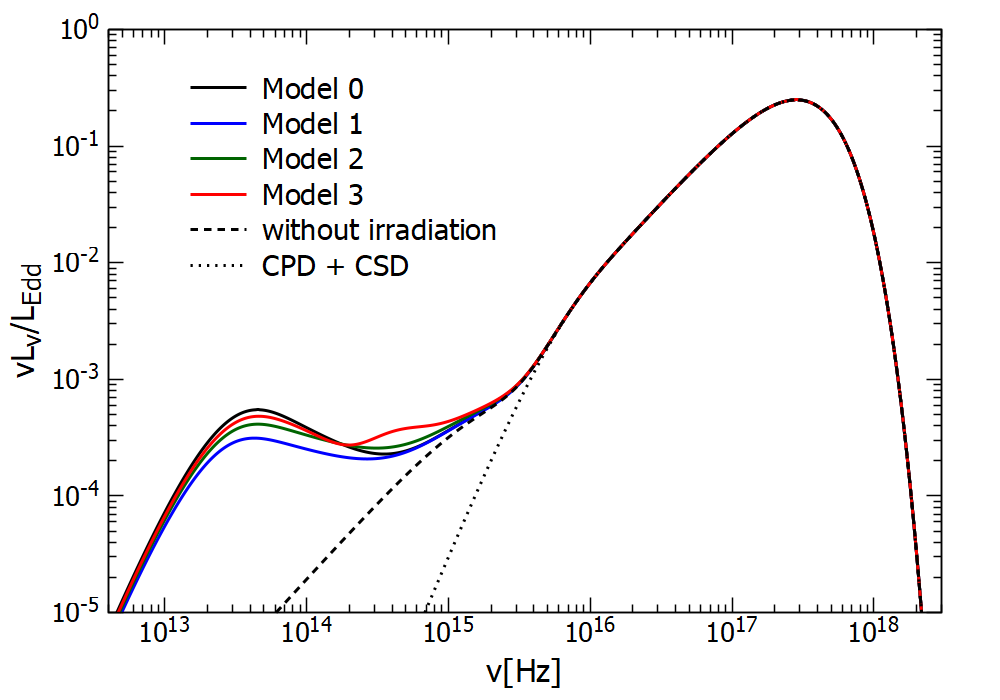}
    \caption{
Comparison of the combined SEDs of the triple disk, consisting of the CPD, the CSD, and the CBD surrounding them. The vertical and horizontal axes represent the normalized luminosity and frequency, respectively, both plotted on a logarithmic scale. For all models, the same CPD and CSD parameters are used, and differences in the triple disk spectra arise solely from the CBD component. The solid black, blue, green, and red lines denote the total SEDs for Models 0, 1, 2, and 3, respectively. The dashed black line shows the triple disk spectrum without irradiation heating, while the dotted black line shows the combined SED from the CPD and CSD only, excluding the CBD contribution. The other CBD parameters are identical to those adopted in panel (a) of Figure~\ref{fig:CBD_mass} for all models.
    }
   \label{fig:triple}
\end{figure*}

Figure~\ref{fig:triple} illustrates how the combined SED of the triple disk system responds to different opacity prescriptions applied to the CBD. Since the CPD and CSD parameters are held fixed across all models, the differences among the spectra reflect only the variation in the CBD treatment. In all cases, the triple disk SED exhibits a double-peaked structure: a high-frequency (X-ray) peak originating from the hot inner regions of the CPD and CSD, and a lower-frequency (IR) peak caused by reprocessed irradiation in the outer CBD, as shown by \cite{2024ApJ...975..141L}. Note that the IR component shows significant model dependence due to the opacity-dependent surface temperature profile of the CBD.

The dashed black curve in Figure~\ref{fig:triple} represents the total SED when the irradiation heating term is omitted from the CBD, thereby isolating the role of viscous heating alone. Compared to the irradiated cases, the IR peak is significantly suppressed, confirming that irradiation is essential for generating the low-frequency excess. The dotted line shows the spectrum produced solely by the CPD and CSD without any CBD contribution. Notably, this curve produces the X-ray peak but lacks the IR component entirely. This comparison highlights the distinct spectral signatures arising from each disk component: the CPD and CSD dominate the high-energy emission, while the CBD is the sole contributor to the IR peak. The deviation between models becomes more prominent in the IR regime, demonstrating the impact of different opacity prescriptions on the CBD's reprocessing efficiency.

\begin{figure*}[htbp]
\centering
\includegraphics[scale = 0.3]{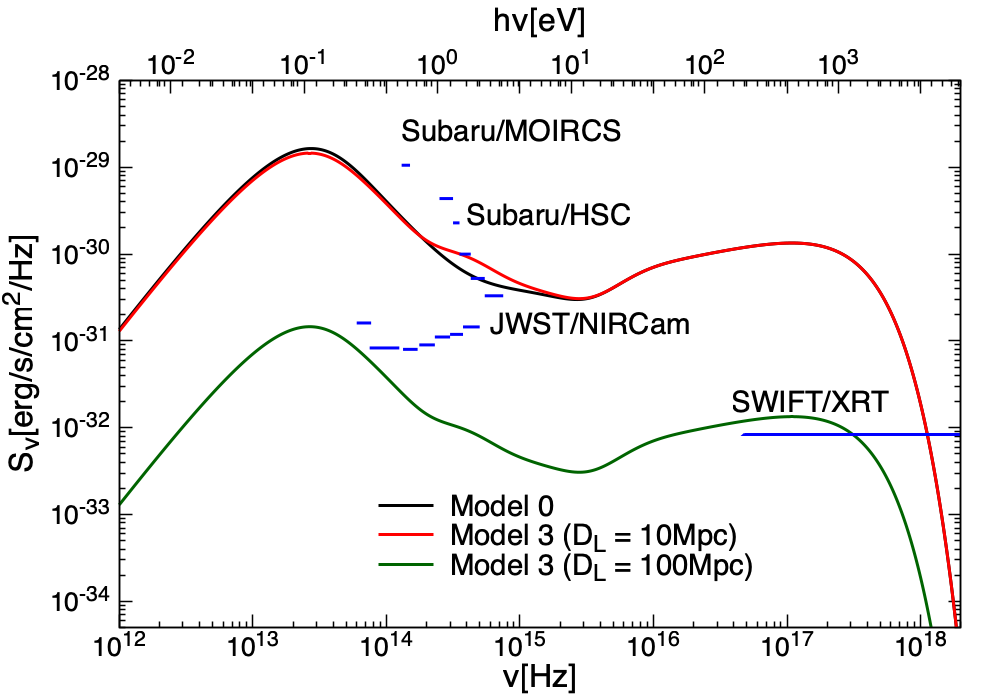}
\caption{
Comparison of Subaru, JWST, and Swift/XRT sensitivity limits with theoretical triple disk spectra at different source distances. The vertical axis shows the flux density (in erg\,s$^{-1}$\,cm$^{-2}$\,Hz$^{-1}$), and the horizontal axis represents the frequency (in Hz); both are plotted on logarithmic scales. All opacity models adopt fiducial parameters of $M = 100\,M_\odot$ and $r_{\rm out} = 10^4\,r_{\rm in}$. The solid black line denotes Model~0, while the solid red and green lines represent Model~3 at distances of $D_{\rm L} = 10\,\mathrm{Mpc}$ and $D_{\rm L} = 100\,\mathrm{Mpc}$, respectively. The dashed horizontal curves indicate the flux sensitivity limits of JWST/NIRCam filters (F444W through F070W), Subaru/MOIRCS and HSC filters ($Y$, $z$, $i2$, $r2$, and $g$), and Swift/XRT in the 0.2--10\,keV X-ray band. This comparison highlights the observability of CBD features in the IR-optical-X-ray bands and demonstrates that, for $D_{\rm L} = 100\,\mathrm{Mpc}$, only the X-ray emission (primarily from the CPD and CSD) is detectable, whereas at $D_{\rm L} = 10\,\mathrm{Mpc}$, the irradiated CBD in Model~3 becomes marginally detectable in the Subaru/HSC $r2$ band.
}
\label{fig:flimit}
\end{figure*}

%
\subsection{Observational implications}
\label{sec:obsimp}
%

To evaluate the detectability of the model spectra, we compare the predicted flux densities with the sensitivity limits of major current telescopes. These include Subaru/HSC and MOIRCS in the optical and near-IR (NIR) bands, JWST/NIRCam in the mid-to-NIR, and Swift/XRT in the X-ray band. Figure~\ref{fig:flimit} shows the triple disk spectra for Models~0 and 3 at source distances of 10~Mpc and 100~Mpc, overlaid with the flux detection thresholds of these instruments. All other model parameters are fixed at their fiducial values ($M = 100\,M_\odot$, $r_{\rm out} = 10^4\,r_{\rm in}$).

At $D = 100\,{\rm Mpc}$, even Model~3, which is our most optically thick model, falls below the detection thresholds in the optical and NIR bands. The only potentially observable signal in this case is the X-ray emission, originating from the two minidisks (CPD and CSD), which is detectable with Swift/XRT. In contrast, at $D = 10\,{\rm Mpc}$, Model~3 exhibits a distinct low-frequency excess due to the irradiated CBD, making it detectable with JWST and also marginally detectable with the Subaru/HSC $r2$ filter. This contrast highlights the role of distance in the detectability of the CBD contribution in the IR-optical regime.

The fact that irradiation-induced CBD features emerge in the optical/NIR only at nearby distances ($\lesssim 10\,{\rm Mpc}$) sets practical constraints on follow-up campaigns. In particular, detecting the IR peak requires deep imaging of relatively nearby sources with multiple filters to disentangle the spectral shape. These features may appear as broad, red excesses in the SED, distinguishable from the steeper blue continuum of minidisk-dominated emission.

Based on this, we propose a targeted follow-up strategy: Upon detection of a bright X-ray flare by Swift/XRT or other X-ray facilities, especially if a distance estimate places the source within $\lesssim 100$~Mpc, NIR imaging should be promptly initiated using telescopes such as Subaru or JWST. If a counterpart is found, a detailed spectro-photometric observation across multiple bands can constrain the SED slope and test for IR flattening predicted by our Model~3. Such multi-band follow-up would be key to identifying reprocessed emission from irradiated CBDs.

%
\subsection{
Gravitational wave detectability
}
\label{sec:gw_detectability}
%
%

Since the binary loses orbital energy and angular momentum through gravitational radiation, the waveform is determined by the orbital parameters. In the quasi-circular regime the GW frequency increases slowly while the strain grows as the orbit shrinks \citep{SathyaprakashSchutz2009}. For our purpose we describe this slow chirp with a frequency-domain treatment that connects the orbital evolution to the observed characteristic strain and the mission-averaged signal-to-noise ratio (SNR).

The characteristic strain under the stationary phase approximation (SPA) is given by (e.g., \citealt{SathyaprakashSchutz2009,2015CQGra..32a5014M})
\begin{equation}
\label{eq:hc}
h_{\rm c}(f) = \frac{8}{\sqrt{5}} T_{\rm obs}^{1/2} \frac{\bigl(G\mathcal{M}\bigr)^{5/3}}{D_{\rm L}c^4} \pi^{2/3} f^{7/6},
\end{equation}
where $\mathcal{M}=\bigr[q/(1+q)^2\bigr]^{3/5}M$ is the chirp mass and the GW frequency is given by
\begin{eqnarray}
f=\frac{2}{P_{\rm orb}}=\frac{1}{\pi}\sqrt{\frac{GM}{a^3}}.
\label{eq:fgw}
\end{eqnarray}
The frequency evolution in the inspiral is then obtained by
\begin{equation}
\label{eq:fdot}
\frac{df}{dt}
= \frac{96}{5}\,\pi^{8/3}
\left(\frac{G\mathcal{M}}{c^{3}}\right)^{5/3} f^{11/3},
\end{equation}
which we use to determine the frequency interval swept during the observing time $T_{\rm obs}$.

\begin{figure}[ht!]
  \centering
    \includegraphics[scale=0.3]{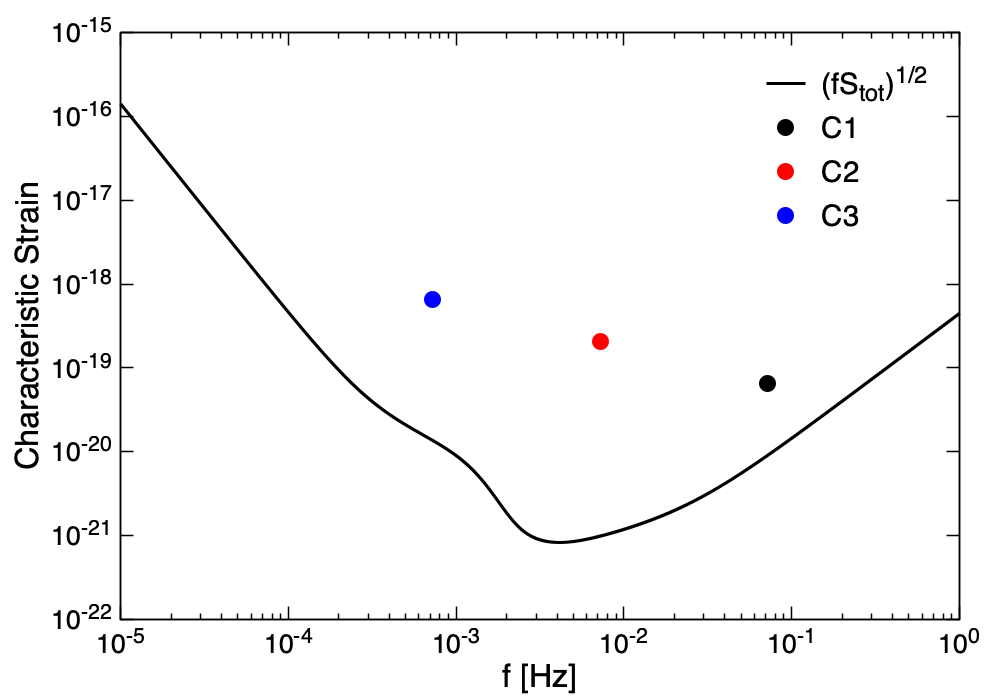}
    \caption{
    Comparison of the LISA sensitivity $S_{\rm tot}^{1/2}(f)$ with the gravitational-wave signals for three different masses: C1 (black), C2 (red), and C3 (blue). The horizontal axis is the GW frequency $f$ in units of Hz, and the vertical axis is the characteristic strain $h_{\rm c}$. Colored points mark $h_{\rm c}$ evaluated at $q=1$, $a=1000\,r_{\rm S}$, and $D_{\rm L} = 10\,\mathrm{Mpc}$, whereas the black line shows $\sqrt{f\,S_{\rm tot}(f)}$. All values assume circular, face-on, equal-mass binaries. It is noted that $S_{\rm tot}(f)$ is the sum of the amplitude spectral densities of the LISA instrument noise $S_{\rm inst}(f)$ and Galactic confusion noise $S_{\rm c}(f)$.
    }
    \label{fig:snr_lisa}
\end{figure}

We compute the mission-averaged SNR using the total LISA noise $S_{\mathrm{tot}}(f)$ of \citet{2019CQGra..36j5011R}, which combines the instrument and Galactic confusion contributions: 
\begin{equation}
\label{eq:snr2}
\bar{\rho}_{\rm SNR}^2 = \int_{f_{0}}^{f_{0}+\Delta{f}} \frac{h_{\rm c}^2(f)}{fS_{\mathrm{tot}}(f)}\,\mathrm{d}f,
\end{equation} 
where $f_{0}$ is the entry frequency at the chosen orbital separation $a=a_0r_{\rm S}$, and $\Delta f \simeq (df/dt)\,T_{\rm obs}$ if the chirp is slow. To complement the EM predictions, we evaluate the LISA detectability of three representative mass cases: C1: $M=10\,M_\odot$, C2: $M=100\,M_\odot$, and C3: $M=1000\,M_\odot$. Throughout this subsection C1--C3 refer to the mass cases and should not be confused with the opacity models 1--3 defined earlier. We fix the semi-major axis at $a=1000\,r_{\rm S}$ and the luminosity distance at $D_{\rm L}=10~\mathrm{Mpc}$ for all three cases. As estimated at the end of Section~\ref{sec:cbdspectra}, the decoupling radius is typically $a_{\rm dec}\sim 70\,r_{\rm S}$ for our fiducial parameters. The $a_0=1000$ cases, therefore, remain in the pre-decoupling regime. The resulting quantities used for the three masses are listed in Table~\ref{tbl:gwmodels}, and their locations relative to the LISA instrument, confusion, and total amplitude spectral densities are shown in Figure~\ref{fig:snr_lisa}.

It is noted from the figure that C2 and C3 are detected with high confidence ($\bar{\rho}_{\rm SNR}\simeq 1.98\times10^2$ and $4.72\times10^1$) under the conventional detection threshold $\bar{\rho}_{\rm SNR}\gtrsim 8$, whereas C1 remains sub-threshold ($\bar{\rho}_{\rm SNR}\simeq 3.64$). For the C2 case, the binary enters the LISA band at a GW frequency of $f_{0} \simeq 7.2~\mathrm{mHz}$, which lies very close to the frequency where the total noise amplitude $S_{\rm tot}^{1/2}(f)$ reaches its minimum. During the mission, the signal then sweeps across a logarithmic frequency interval of $\Delta\ln f \simeq 7.7\times10^{-2}$, so it spends a significant time in this most sensitive part of the band. This combination of near–optimal frequency and non-negligible bandwidth yields the largest SNR among the three cases. C1 lies on the high-frequency rise where $S_{\rm inst}^{1/2}$ increases, whereas C3 lies in the sub-mHz region dominated by the confusion foreground; both suffer reduced SNR relative to C2.

Since $h_c \propto D_{\rm L}^{-1}$, the SNR scales as $\bar{\rho}_{\rm SNR}\propto D_{\rm L}^{-1}$, at $D_{\rm L} = 100$~Mpc the three cases give roughly $\bar{\rho}_{\rm SNR} \approx 0.36$ (C1), $19.8$ (C2), and $4.7$ (C3). For fixed $a/r_{\rm S}$, the available $\Delta\ln f$ decreases approximately with increasing $M$ over a finite observing time, which partly offsets the intrinsic rise of $h_c$ with mass.

In our calculation of $\bar{\rho}_{\rm SNR}$ we use the SPA and integrate the signal power along the entire inspiral frequency track. For the noise, we adopt the total LISA power spectral density $S_{\rm tot}(f)$, which includes both the instrumental noise and the Galactic confusion noise. Even if we vary the assumed observing time within a plausible range of a few years, or modify the level of confusion–noise subtraction within realistic limits, the resulting SNRs change only by factors of order unity (typically less than a factor of a few). The qualitative ranking of the three cases, $C2 > C3 \gg C1$, remains unchanged.

Among the three cases, C2 is the most promising for joint GW--EM observations:
it is the loudest in LISA and, in our irradiated-CBD framework, it exhibits a strong IR/optical reprocessing signature. 
For very nearby events, C1 becomes detectable only at distances of a few Mpc, while C3 remains detectable within $\lesssim 10\,\mathrm{Mpc}$, enabling direct tests of the opacity-dependent CBD spectra presented in Sections~\ref{sec:cbdspectra}--\ref{sec:obsimp}

%
%
\begin{table*} [htbp]
    \renewcommand{\arraystretch}{1.0}
    \centering
        \begin{tabular}{c|c|c|c|c|c}
            \hline
             Case 
             & Binary mass ($M$)
             & GW frequency ($f_{}$)
             & $\Delta\ln f$ 
             & Characteristic strain ($h_{\rm c}$)
             & SNR ($\mathcal{\bar{\rho}}_{\rm SNR}$)\\
            & $M_\odot$
            & ${\rm Hz}$
            & $-$
            & $-$
            & $-$
            \\
            \hline
            C1 & $10$ & $7.2\times10^{-2}$ & $7.7\times10^{-1}$ & $6.5\times10^{-20}$ & $3.6$\\
            C2 & $100$ & $7.2\times10^{-3}$ & $7.7\times10^{-2}$ &$2.0\times10^{-19}$ & $198$\\
            C3 & $1000$ & $7.2\times10^{-4}$ & $7.7\times10^{-3}$ & $6.5\times10^{-19}$ & $47$\\ 
            \hline
        \end{tabular}
    \caption{
   GW observables and SNR for the three BBHs with different masses, evaluated at a common orbital separation $a_0=1000$. Columns list the model label, binary mass $M$, GW frequency $f_{}$, characteristic strain $h_c$ (quadrupole estimate at a common luminosity distance $D_{\rm L} = 10\,\mathrm{Mpc}$, and SNR $\bar{\rho}_{\rm SNR}$ for LISA. All binaries are assumed circular, face-on, and equal-mass. 
   }
    \label{tbl:gwmodels}
\end{table*}

%
\section{Discussion}
\label{sec:dis}
%

\begin{figure}
\centering
\includegraphics[scale=0.3]{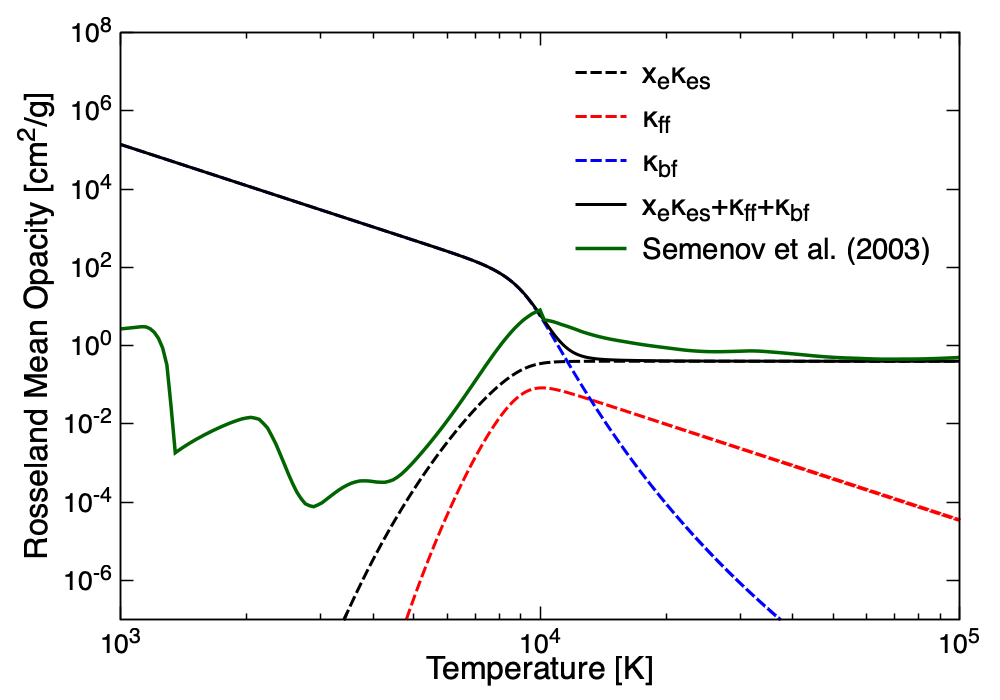}
\caption{
Comparison of analytic hydrogen opacities and tabulated Rosseland-mean opacities at a fixed density of $\rho=10^{-10}\,{\rm g\,cm^{-3}}$. The dashed black, red, and blue lines represent the contributions from electron scattering, free-free absorption, and bound-free absorption for hydrogen gas, respectively. The solid black line represents the total analytical opacity, calculated as $x_e\kappa_{\rm es}+\kappa_{\rm ff} + \kappa_{\rm bf}$, where the ionization fraction $x_e$ is determined self-consistently from the Saha equation. The solid green line indicates the tabulated opacity from the Opacity Project \citep{2005MNRAS.362L...1S}, an international collaboration to calculate the extensive atomic data required to estimate stellar envelope opacities, and from \citep{2003A&A...410..611S}, including dust opacity based on normal silicate grains under the assumption of homogeneous spheres.
}
\label{fig:opacities}
\end{figure}

This section discusses several key caveats of our models. First, we examine how realistic opacity prescriptions affect the emergent spectrum from an irradiated CBD, focusing on discrepancies between our opacity models and more detailed opacities available in tabulated forms. Second, we assess the detectability of GWs emitted from such systems by current and future GW observatories, based on the binary orbital parameters.

%
\subsection{Opacity Prescriptions and Comparison with Tabulated Opacities}
\label{sec:5.1}
%

We now compare our analytic hydrogen opacities with widely used tabulated opacity tables. Figure~\ref{fig:opacities} compares the Rosseland mean opacity computed from our opacity models, which include free–free absorption, bound–free absorption, and electron scattering processes, with the tabulated values from the Opacity Project \citep{2005MNRAS.362L...1S}, which provides extensive atomic data for stellar envelopes, and from \citet{2003A&A...410..611S}, which are widely used in protoplanetary disk modeling. In addition to tabulated opacities, protostellar disk studies commonly use piecewise, temperature-dependent fits for the low-temperature Rosseland-mean opacity (e.g., \citealt{BellLin1994,BellEtAl1997}). Our model employs cross-section formulas and ionization fractions derived from the Saha equation, assuming a pure hydrogen composition at a fixed density of $\rho = 10^{-10}~{\rm g\,cm^{-3}}$. The comparison reveals significant discrepancies, particularly in the temperature range $T \sim 10^3$–$10^4~{\rm K}$. In this range, the tabulated Rosseland opacities are several orders of magnitude lower than those predicted by our analytical prescription that includes bound–free absorption. This difference arises because the tabulated opacities incorporate more realistic astrophysical conditions, such as grain sublimation, grain growth, and composition-dependent absorption features, all of which are absent from our simplified hydrogen-only framework.

These differences in opacity have direct consequences for the emergent spectrum from irradiated CBDs. In the temperature range of $T \sim 10^3$–$10^4$ K, the reduced opacity in the tabulated models is expected to enhance radiative cooling, so that energy is transported more efficiently to the disk surface and the surface temperature becomes higher. Consequently, the overall spectral intensity, particularly in the far-IR and submillimeter bands dominated by reprocessed irradiation, increases.

In the supermassive regime, equation (\ref{eq:qirr_app}) implies that the irradiation–dominated heating rate scales as $Q_{\rm irr}\propto M^{-21/19}$. Combined with $Q_{\rm irr}=2\sigma T_{\rm s}^4$, this yields $T_{\rm s}\propto M^{-21/76}$. Accordingly, extending our fiducial $100\,M_\odot$ model to $M\sim10^6$–$10^8\,M_\odot$ would reduce the outer surface temperature by more than an order of magnitude, from $\sim 10^3$–$10^4$\,K to $T_{\rm s}\lesssim$ a few $\times 10^2$\,K. At such temperatures, dust and metal opacities dominate over hydrogen bound–free and free–free opacity, and full tabulated Rosseland means (e.g. \citealt{2003A&A...410..611S,2005ApJ...623..585F,2005MNRAS.362L...1S}) are required for reliable spectra. A self-consistent treatment of the circumbinary SEDs of SMBH binaries therefore demands replacing the analytic Kramers prescriptions employed here with composition-dependent opacity tables, and we defer this extension to future work, restricting the present quantitative SED calculations to stellar- and intermediate-mass BBHs while summarizing only the SMBH mass-scaling.

As a future direction, it will be essential to incorporate opacity tables (e.g., \citealt{2003A&A...410..611S,2005MNRAS.362L...1S}) directly into irradiated disk models, enabling systematic computation and comparison of CBD spectra under more realistic astrophysical conditions. This will provide a more robust framework for interpreting multiwavelength observations and constraining the physical parameters of circumbinary environments.

%
\subsection{Orbital Decay, Circumbinary Disk 
Decoupling, and GW Detection Across Mass Scales}
%

In Section~\ref{sec:gw_detectability}, we evaluated the LISA SNRs for three mass cases (C1–C3) at an orbital separation of $a = 1000\,r_{\rm S}$. Figures~\ref{fig:flimit} and Table~\ref{tbl:gwmodels} show that, in particular, models C2 and C3 reach high SNRs at a luminosity distance of $D_{\rm L} \simeq 10\,{\rm Mpc}$. Using equation~(\ref{eq:fgw}), the corresponding GW frequencies lie in the range $f_{} \sim 7.2\times10^{-2}$–$7.2\times10^{-4}\,{\rm Hz}$ for C1–C3. These values are several orders of magnitude below the most sensitive band ($10$–$10^{3}\,{\rm Hz}$) of ground-based interferometers such as LIGO \citep{1992Sci...256..325A}, Virgo \citep{2015CQGra..32b4001A}, and KAGRA \citep{2013PhRvD..88d3007A}, and fall instead in the millihertz band targeted by LISA \citep{2019CQGra..36j5011R}. Extending the same scaling relation to SMBH binaries with total masses $M \sim 10^{5}$–$10^{9}\,M_{\odot}$, the GW frequency at $a = 1000\,r_{\rm S}$ decreases in proportion to $M^{-1}$, spanning $f_{}(a) \sim 10^{-6}$–$10^{-10}\,{\rm Hz}$. For example, we obtain $f_{}(a) \sim 10^{-7}\,{\rm Hz}$ for $M \sim 10^{6}\,M_{\odot}$ and $f_{}(a) \sim 10^{-8}$–$10^{-9}\,{\rm Hz}$ for $M \sim 10^{7}$–$10^{8}\,M_{\odot}$. Systems in this intermediate mass range fall squarely in the nanohertz band probed by pulsar timing arrays (PTAs) \citep[e.g.,][]{Agazie2023,2019MNRAS.485.1579K}, while $10^{5}\,M_{\odot}$ and $10^{9}\,M_{\odot}$ binaries occupy the adjoining microhertz and sub-nanohertz regimes. SMBH binaries that have shrunk to $a \sim 5$–$50\,r_{\rm S}$ just before coalescence then sweep through the milli- to microhertz band, reaching $f_{} \sim 10^{-5}$–$10^{-2}\,{\rm Hz}$ depending on mass and separation, and become prime LISA targets \citep[e.g.,][]{2009ApJ...700.1952H,2019CQGra..36j5011R}. Multiband GW observations of SMBH binaries are therefore expected to combine PTAs and LISA, whereas the quasi-steady triple-disk phase at $a=1000\,r_{\rm S}$ studied here is not directly accessible to current ground-based detectors.

Let us briefly discuss dynamical decoupling for all these mass scales. Following equation~(\ref{eq:adec}), the decoupling radius in units of Schwarzschild radius is estimated to be $a_{\rm dec}/r_{\rm S}\sim70$ for the fixed dimensionless parameters ($q$, $C_{\rm gap}$, $\alpha_{\rm SS}$, and $Y_{\rm in}$) and thus is independent of the binary mass $M$ (although $a_{\rm dec}\propto M$ in physical units). For our fiducial parameters, both $a=100\,r_{\rm S}$ and $a=1000\,r_{\rm S}$ lie in the pre-decoupling regime, even when the system is scaled to SMBH binaries. Once the binary shrinks to $a\lesssim a_{\rm dec}$, the inflow into the central cavity is suppressed, the structure and luminosity of any minidisks around each black hole change significantly, and post-merger inflow of the CBD toward the remnant black hole may produce an afterglow-like rebrightening \citep{MilosavljevicPhinney2005}. EM emission associated with the LIGO band and the high-frequency end of the LISA band therefore requires a different class of time-dependent disk models that explicitly include CBD decoupling and evolving accretion (e.g., \citealt{armitage_accretion_2002,2023ApJ...949L..30D}), whereas the framework developed in this work is mainly designed to describe the earlier, strongly coupled phase.

%
\subsection{Light-curve Periodicity, 
Time-averaged SED, and Disk Structure Assumptions}
%

Early analytic work on CBD-assisted binary evolution and its electromagnetic signatures (e.g., \citealt{armitage_accretion_2002,ArmitageNatarajan2005,MilosavljevicPhinney2005,2009ApJ...700.1952H,2009PASJ...61...65H})
motivated hydrodynamic studies of circumbinary accretion, which established that streams penetrating the central cavity can feed minidisks in a strongly time-dependent manner (e.g., \citealt{hayasaki_binary_2007,hayasaki_supermassive_2008,2008ApJ...672...83M,2009MNRAS.393.1423C}). Self-consistent azimuthally averaged steady-state solutions for strongly perturbed CBDs were later derived analytically \citep{2012MNRAS.427.2660K,2012MNRAS.427.2680K}. Subsequent 2D/3D hydrodynamic and MHD simulations and parameter surveys have clarified that the variability is often concentrated near characteristic periods of $\sim 5\,P_{\rm orb}$ for near-circular equal-mass binaries and $\sim P_{\rm orb}$ for eccentric systems, with additional beat-frequency components and pronounced minidisk asymmetries depending on mass ratio, disk thickness, thermodynamics, viscosity, and magnetic flux
(e.g., \citealt{2013MNRAS.436.2997D,Shi2012,2015ApJ...807..131S,MunozLai2016,Miranda2017,Moody2019,DuffellEtAl2020,2018ApJ...853L..17B,Bowen2019,DittmannRyan2022,MostWang2024,LaiMunoz2023}).
Our focus, however, is the time-averaged SED obtained from this 1D, steady, orbit-averaged CBD model. At radii relevant for reprocessing (e.g., $r \sim 10\,a$), the thermal timescale $t_{\rm th} \simeq (\alpha \Omega)^{-1}$ is $\gg P_{\rm orb}$, and as a result, short-period fluctuations are strongly smoothed. Hence, binary imprints appear primarily through irradiation geometry and inner truncation, shaping the global SED rather than producing sharp periodic features. A fully time-dependent treatment that follows modulated accretion rates and cavity evolution is left for future work.

In this work, we have adopted the assumption that the two black holes have equal masses so that our prescription (\ref{eq:mdots12}) implies equal accretion rates onto the two minidisks. However, recent parameter-space surveys show in detail how accretion-rate splitting depends on the mass ratio, disk thickness, and viscosity, providing guidance beyond the equal-splitting assumption (e.g., \citealt{2024ApJ...967...12D}). These prescriptions and their variants are already employed in population and detectability studies of binary SMBHs \citep{2019MNRAS.485.1579K}. Future studies will address the incorporation of these parameter dependencies into our framework.

Finally, we note that the structural equations employ gas pressure only (equations (\ref{eq:hydrostaticeq}) and (\ref{eq:soundspeed})–(\ref{eq:mid-plane-temp})); radiation and magnetic pressure are omitted. For hotter or near-Eddington disks where radiation pressure can dominate, the vertical structure and $T_{\rm s}$ may change. Extending the framework to include radiation pressure is a natural next step.

%
\section{Conclusions}
\label{sec:conclusion}
%

We have investigated the effect of opacity prescriptions on the thermal and spectral properties of irradiated CBDs surrounding BBH systems. Our models incorporated temperature- and density-dependent hydrogen opacities, specifically considering electron scattering, free-free absorption, and bound-free absorption. By numerically solving the local energy balance equation with self-consistent outer boundary conditions, we analyzed how different opacity prescriptions shape the temperature profiles and resultant spectra of the CBD. Our main findings are summarized as follows:

\begin{enumerate}
\item The CBD surface temperature profiles remain largely consistent across different opacity models in the regions dominated by viscous heating. However, significant differences occur in the midplane temperature profiles due to the substantial influence of disk opacity.

\item Opacity effects predominantly manifest in the lower-frequency regime of the CBD spectrum, where external irradiation from minidisks dominates heating. These effects are most pronounced in the IR and optical bands.
\item Inclusion of analytical hydrogen absorption opacities notably flattens the continuum emission in the IR-optical range, creating an additional spectral hump distinct from the standard double-peaked profile identified in the earlier simpler models \citep{2024ApJ...975..141L}.
\item
Despite the difference in the opacity prescriptions, the combined spectrum of the triple disk system, which consists of the CBD and two minidisks, retains a characteristic double-peaked structure. The high-frequency X-ray peak primarily arises from the inner regions of the minidisks, while the low-frequency IR peak results from irradiation reprocessing in the CBD.
\item Observational feasibility analysis demonstrates that current instruments like Subaru/HSC, JWST/NIRCam, and Swift/XRT provide adequate sensitivity for distinguishing among our opacity models, particularly at distances less than $\sim10$~Mpc. Hence, coordinated multi-band follow-up observations following X-ray flares could effectively test our model predictions and discern the physical conditions in irradiated CBDs.

\item 
The CBD spectral shape is primarily controlled by $\xi_{\rm out}$, $M$, and $a$. In particular, $\xi_{\rm out}$ sets the emitting area of the irradiation-powered IR bump and thus its low-frequency extent, while $M$ and $a_0$ largely determine the global and inner-disk frequency scales, respectively.

\item 
For binaries at $a = 1000\,r_{\rm S}$, the GW signal lies in the LISA band and the detectability depends strongly on mass: the C2 ($M=100\,M_\odot$) and C3 ($M=1000\,M_\odot$) cases are detectable with high SNRs at $D_{\rm L}\simeq 10\,{\rm Mpc}$, whereas the C1 ($M=10\,M_\odot$) case remains sub-threshold except for very nearby events. This favors joint EM--GW follow-up for intermediate-mass systems.

\item 
When the irradiation-heated CBD scalings are extrapolated to supermassive BBH masses, the outer surface temperature decreases to $T_{\rm s}\lesssim {\rm a\ few}\times 10^{2}\,{\rm K}$, where dust and metal opacities dominate. Quantitative predictions for the CBDs therefore require composition-dependent tabulated Rosseland-mean opacities beyond the hydrogen-only framework adopted here.
\end{enumerate}

Our results underscore the critical importance of accurately modeling opacity in astrophysical disk systems, highlighting the necessity for incorporating more realistic opacity prescriptions in future theoretical studies. This advancement will significantly enhance our ability to interpret EM signals associated with GW events from BBHs approaching coalescence.

%
\section*{Acknowledgments}
%
We thank the referee for the constructive and generous suggestions that have improved the paper.
S.B. acknowledges support from the Basic Science Research Program through the National Research Foundation of Korea (NRF), funded by the Ministry of Education (grant RS-2024-00460704).
This work was supported by the National Research Foundation of Korea (NRF) grant funded by the Korea government (MSIT) (2020R1A2C1007219 and RS-2025-23323627).

This work was also conducted during the research year of Chungbuk National University in 2021. This work was partially supported by NSF grant PHY-2309135 to the Kavli Institute for Theoretical Physics (KITP), and by the Grant-in-Aid for Scientific Research from MEXT/JSPS of Japan (grant JP21K03619 to A.T.O.).

\appendix

%
\section{Analytical solutions for $\beta\neq0$}
\label{app:anasol}
%

%
\subsection{Optical depth prescriptions}
%

In this section, we describe the optical depth prescriptions. 
Equations~(\ref{eq:tau1})-(\ref{eq:tau3}), which show the optical depths as a function of $Y$ and $\xi$, are rewritten as
\begin{eqnarray}
    \tau_1
    &=&
    \frac{1}{2}\kappa_{\rm es}\Sigma
    = 
    \tau_{10}
    \frac{1}{Y^{2}\xi^{1/2}},
    \label{eq:tau1a}
        \\
    \tau_2
    &=&
    \frac{1}{2}\sqrt{\kappa_{\rm es}\kappa_{\rm ff}}\;\Sigma
    = \tau_{20}\, \frac{\xi^{1/2}}{Y^{7}},
    \label{eq:tau2a}
\\
    \tau_3
    &=&
        \frac{1}{2}\sqrt{\kappa_{\rm a}\!\left(\kappa_{\rm a}+\kappa_{\rm es}\right)}\;\Sigma
    =\frac{1}{2}\kappa_3\Sigma
    =
\tau_{30}\frac{\xi^{3/2}}{Y^{12}}
\,
\Biggr[1+\tau_{31}\frac{Y^{10}}{\xi^2}\Biggr]^{1/2}
\approx
\tau_{30}\frac{\xi^{3/2}}{Y^{12}}
+
\frac{\tau_{10}}{2}\frac{1}{\xi^{1/2}Y^2}
\approx
\tau_{30}\frac{\xi^{3/2}}{Y^{12}}
,
    \label{eq:tau3a}
\end{eqnarray}
where $x_e=1$ for $\tau_{20}$ and $x_e=0$ for $\tau_{30}$ and $\tau_{31}$ are adopted so that
\begin{eqnarray}
\tau_{10}
    &=&
    \frac{1}{\sqrt{2}}\frac{\dot{m}}{\alpha_{\rm SS}}
    \frac{1}{(C_{\rm gap}a_0)^{1/2}},
\label{eq:tau10}
    \\
\tau_{20}
    &=&
        \frac{1}{2^{3/4}}\left(\frac{\dot{m}}{\alpha_{\rm SS}}\right)^{3/2}\frac{1}{(C_{\rm gap}a_0)^{5/4}}   
    \left(\frac{\kappa_{\rm ff,0}}{r_{\rm S}}\right)^{1/2}
    \frac{x_e}{\kappa_{\rm es}} 
\left(
\frac{1+x_e}{T_0}
\right)^{7/4}
=
     2
    \left(\frac{\dot{m}}{\alpha_{\rm SS}}\right)^{3/2}\frac{1}{(C_{\rm gap}a_0)^{5/4}}   
    \left[
    \left(
    \frac{\kappa_{\rm ff,0}}{ \kappa_{\rm es}^2r_{\rm S}}
    \right)
    T_0^{-7/2}
    \right]^{1/2},
    \label{eq:tau20}
     \\
\tau_{30}
    &=&
    \frac{1}{2}\left(\frac{\dot{m}}{\alpha_{\rm SS}}\right)^2\frac{1}{(C_{\rm gap}a_0)^{2}}
    \frac{\kappa_{\rm ff,0}}{\kappa_{\rm es}^2}
   \frac{1}{r_{\rm S}}
\left(
\frac{1+x_e}{T_0}
\right)^{7/2}
\left[
x_e^2 + (1-x_e)
\frac{\kappa_{\rm bf,0}}{\kappa_{\rm ff,0}}
\right]
=
    \frac{1}{2}\left(\frac{\dot{m}}{\alpha_{\rm SS}}\right)^2\frac{1}{(C_{\rm gap}a_0)^{2}}
    \frac{\kappa_{\rm bf,0}}{\kappa_{\rm es}^2r_{\rm S}}{T_0}^{-7/2},
    \label{eq:tau30}
    \\
    \tau_{31}
    &=&
    \sqrt{2}\frac{\alpha_{\rm SS}}{\dot{m}}(C_{\rm gap}a_0)^{3/2}
    \frac{\kappa_{\rm es}^2r_{\rm S}}{\kappa_{\rm ff,0}}
    \left(
\frac{T_0}{1+x_e}
\right)^{7/2}
    \left[
\frac{1}{
x_e^2 + (1-x_e)
(\kappa_{\rm bf,0}/\kappa_{\rm ff,0})
}
\right]
=
    \sqrt{2}\frac{\alpha_{\rm SS}}{\dot{m}}(C_{\rm gap}a_0)^{3/2}
    \frac{\kappa_{\rm es}^2r_{\rm S}}{\kappa_{\rm bf,0}}
    T_0^{7/2}. 
    \nonumber
\end{eqnarray}

%
\subsection{Analytical solutions}
%

For each model, equation~(\ref{eq:dydx}) is expressed as
\begin{eqnarray}
    \left[
\xi^3+\frac{\beta}{2}\xi 
\right]
    \frac{dY}{d\xi}
    &=&
    \left[
    \frac{1}{3}\frac{\alpha}{\tau_{10}}
    Y^{10}\xi^{1/2}
    +\beta\,Y
    \right],
\label{eq:tau1deq} 
\\
       \left[
\xi^3+\frac{\beta}{2}\xi 
\right]
    \frac{dY}{d\xi}
    &=&
    \left[
    \frac{1}{3}\frac{\alpha}{\tau_{20}}
    \frac{Y^{15}}{\xi^{1/2}}
    +\beta\,Y
    \right], 
\label{eq:tau2deq}
\\
\left[
\xi^3+\frac{\beta}{2}\xi 
\right]
    \frac{dY}{d\xi}
    &=&
    \left[
    \frac{16}{3}\frac{\alpha}{\tau_{30}}
    \frac{Y^{20}}{\xi^{3/2}}
    +\beta\,Y
    \right], 
\label{eq:tau3deq}
\end{eqnarray}
where $x_e=1$ is adopted for Models~ 1 and 2, while $x_e=0$ is adopted for Model 3. Now we solve these differential equations by the perturbation method with $\beta$, which is much smaller than $1$. Considering that the terms with $\beta$ in the above equations are much smaller than those without $\beta$, we can expand $Y$ for $\beta\ll1$ as
\begin{eqnarray}
Y=Y_0(\xi)+\beta{Y_1(\xi)}+\beta^2{Y_2(\xi)}+\cdots.
\label{eq:psol}
\end{eqnarray}
Substituting equation~(\ref{eq:psol}) into equations~(\ref{eq:tau1deq})-(\ref{eq:tau3deq}), we obtain the following equations, which are approximated at the first-order of $\beta$, as
\begin{eqnarray}
    \xi^3\frac{dY_0}{d\xi} 
    + 
    \beta
    \left(\frac{\xi}{2}\frac{dY_0}{d\xi}+\xi^3\frac{dY_1}{d\xi}
    \right)
    &=&
\frac{1}{3}\frac{\alpha}{\tau_{10}}
    Y_0^{10}\xi^{1/2}
+
\beta
\left(
\frac{10}{3}\frac{\alpha}{\tau_{10}}Y_0^9Y_1\xi^{1/2}
+
Y_0
\right),
\\
\xi^3\frac{dY_0}{d\xi} 
    + 
    \beta
    \left(\frac{\xi}{2}\frac{dY_0}{d\xi}+\xi^3\frac{dY_1}{d\xi}
    \right)
    &=&
\frac{1}{3}\frac{\alpha}{\tau_{20}}
    Y_0^{15}\xi^{-1/2}
+
\beta
\left(
5\frac{\alpha}{\tau_{20}}Y_0^{14}Y_1\xi^{-1/2}
+
Y_0
\right),
\\
\xi^3\frac{dY_0}{d\xi} 
    + 
    \beta
    \left(\frac{\xi}{2}\frac{dY_0}{d\xi}+\xi^3\frac{dY_1}{d\xi}
    \right)
    &=&
\frac{16}{3}\frac{\alpha}{\tau_{30}}
    Y_0^{20}\xi^{-3/2}
+
\beta
\left(
\frac{320}{3}
\frac{\alpha}{\tau_{30}}Y_0^{19}Y_1\xi^{-3/2}
+
Y_0
\right).
\end{eqnarray}
For the terms with the 0th order of $\beta$, we obtain the following identities:
\begin{eqnarray}
    \frac{dY_0}{d\xi} 
    &=&
    \frac{1}{3}\frac{\alpha}{\tau_{10}}
    Y_0^{10}\xi^{-5/2},
      \label{eq:dy0dx1}
\\
    \frac{dY_0}{d\xi} 
    &=&
    \frac{1}{3}\frac{\alpha}{\tau_{20}}
    Y_0^{15}\xi^{-7/2},
    \label{eq:dy0dx2}
\\
    \frac{dY_0}{d\xi} 
    &=&
    \frac{16}{3}\frac{\alpha}{\tau_{30}}
    Y_0^{20}\xi^{-9/2},
    \label{eq:dy0dx3}
\end{eqnarray}
yielding the power-law solutions:
\begin{eqnarray}
      Y_0
      &=&
      \left(
      \frac{\tau_{10}}{2\alpha}
      \right)^{1/9}
      \xi^{1/6},
      \label{eq:y01}
\\
      Y_0
      &=&
      \left(
      \frac{15}{28}
      \frac{\tau_{20}}{\alpha}
      \right)^{1/14}
      \xi^{5/28},
      \label{eq:y02}
\\
      Y_0
      &=&
      \left(
      \frac{21}{608}
      \frac{\tau_{30}}{\alpha}
      \right)^{1/19}
      \xi^{7/38}.
      \label{eq:y03} 
\end{eqnarray}

For the terms with the $1$st order of $\beta$, we get the following identities:
\begin{eqnarray}
\frac{\xi}{2}\frac{dY_0}{d\xi}+\xi^3\frac{dY_1}{d\xi}
    &=&
\frac{10}{3}\frac{\alpha}{\tau_{10}}Y_0^9Y_1\xi^{1/2}
+
Y_0,
\label{eq:dy1dx1}
\\
\frac{\xi}{2}\frac{dY_0}{d\xi}+\xi^3\frac{dY_1}{d\xi}
    &=&
5\frac{\alpha}{\tau_{20}}Y_0^{14}Y_1\xi^{-1/2}
+
Y_0,
\label{eq:dy1dx2}
\\
    \frac{\xi}{2}\frac{dY_0}{d\xi}+\xi^3\frac{dY_1}{d\xi}
    &=&
\frac{320}{3}
\frac{\alpha}{\tau_{30}}
Y_0^{19}Y_1\xi^{-3/2}
+
Y_0.
\label{eq:dy1dx3}
\end{eqnarray}
Substituting equations~(\ref{eq:dy0dx1})-(\ref{eq:y03}) into equations~(\ref{eq:dy1dx1})-(\ref{eq:dy1dx3}) yields the following solutions: 
\begin{eqnarray}
Y_1
&=&
-\frac{11}{42}      
\left(
\frac{\tau_{10}}{2\alpha}
\right)^{1/9}
\xi^{-11/6},
\label{eq:y11}
\\
Y_1
&=&
    -\frac{17}{84}
     \left(
      \frac{15}{28}
      \frac{\tau_{20}}{\alpha}
      \right)^{1/14}
      \xi^{-51/28},
\label{eq:y12}
\\
Y_1
&=&
- \frac{69}{418}
 \left(
 \frac{21}{608}
 \frac{\tau_{30}}{\alpha}
 \right)^{1/19}
 \xi^{-69/38}.
 \label{eq:y13}
\end{eqnarray}

From equations~(\ref{eq:y01}) and (\ref{eq:y13}), we obtain the approximate solutions as
\begin{eqnarray}
    Y(\xi; \tau_{10})&=&Y_0+\beta\,Y_1
    =
    \left(
\frac{\tau_{10}}{2\alpha}
\right)^{1/9}
\xi^{1/6}
\left[
1-\frac{11}{42}\frac{\beta}{\xi^2}
\right],
\label{eq:anaytau1}
\\
    Y(\xi; \tau_{20})
    &=&Y_0+\beta\,Y_1
    =
    \left(
\frac{15}{28}\frac{\tau_{20}}{\alpha}
\right)^{1/14}
\xi^{5/28}
\left[
1-\frac{17}{84}\frac{\beta}{\xi^2}
\right],
\label{eq:anaytau2}
\\
    Y(\xi;\tau_{30})
    &=&Y_0+\beta\,Y_1
    =
    \left(
    \frac{21}{608}
\frac{\tau_{30}}{\alpha}
\right)^{1/19}
\xi^{7/38}
\left[
1
-
\frac{69}{418}\frac{\beta}{\xi^2}
\right].
\label{eq:anaytau3}
\end{eqnarray}
with fixed values of $\alpha$ and $\beta$.

%
\bibliography{sb}{}
\bibliographystyle{aasjournal}
%

\end{document}